\documentclass[ aps, twocolumn, nofootinbib, showkeys, notitlepage, superscriptaddress]{revtex4-2}
\usepackage{natbib}
\usepackage{lipsum}
\usepackage{graphicx}
\usepackage{dcolumn}
\usepackage{bm}
\usepackage{amsmath}
\usepackage{float}
\usepackage{multirow}
\usepackage{slashed}
\usepackage{xcolor}
\usepackage{physics}
\usepackage{multirow}
\usepackage{gensymb}
\usepackage[colorlinks=true, pdfstartview=FitV, bookmarks=true, bookmarksnumbered=true, breaklinks]{hyperref}
\usepackage{mathtools,braket}
\usepackage{lipsum}  
\usepackage{color}
\definecolor{blue}{rgb}{0.0, 0.0, 1.0}
\definecolor{red}{rgb}{1.0, 0.0, 0.0}
\definecolor{royalblue}{rgb}{0.0, 0.14, 0.4}
\hypersetup{linkcolor=blue, citecolor=blue, urlcolor=blue}

\def\orcid#1{\kern .08em\href{https://orcid.org/#1}{\includegraphics[keepaspectratio,width=0.7em]{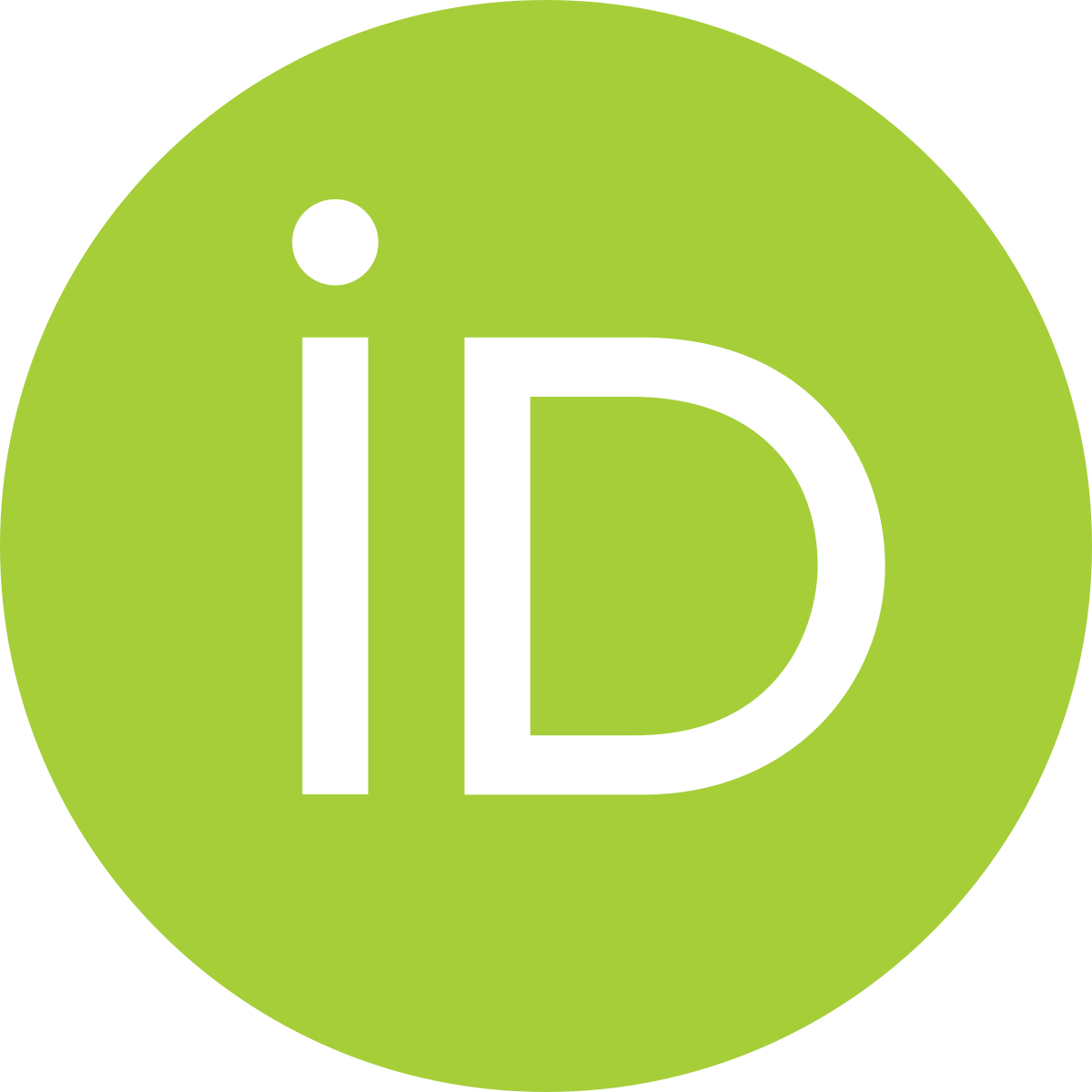}}}

\begin{document}

\title{Structure of heavy quarkonia in a strong magnetic field}

\author{Ahmad Jafar Arifi\orcid{0000-0002-9530-8993}}
\email{aj.arifi01@gmail.com}
\affiliation{Advanced Science Research Center, Japan Atomic Energy Agency, Tokai 319-1195, Japan}
\affiliation{Research Center for Nuclear Physics, The University of Osaka, Ibaraki, Osaka 567-0047, Japan}

\author{Kei Suzuki\orcid{0000-0002-8746-4064}}
\email{k.suzuki.2010@th.phys.titech.ac.jp}
\affiliation{Advanced Science Research Center, Japan Atomic Energy Agency, Tokai 319-1195, Japan}

\date{\today}

%=========================================================
\begin{abstract}
%=========================================================

We investigate the structural modifications of heavy quarkonia in the presence of strong magnetic fields using a constituent quark model. By incorporating the effects of spin mixing and quark Landau levels, we employ a nonrelativistic Hamiltonian that captures the essential features of quark dynamics in a magnetic field. The two-body Schrödinger equation is solved using the cylindrical Gaussian expansion method, which respects the cylindrical symmetry induced by a magnetic field. 
We extract the corresponding light-front wave function (LFWF) densities and analyze their transverse and longitudinal structures, revealing characteristic features such as transverse momentum broadening.
While the longitudinal structure is only slightly modified within the nonrelativistic Hamiltonian, 
we discuss some corrections that can significantly affect its longitudinal structure.
Furthermore, we discuss the structure modifications of excited states and find notable changes in the LFWF densities, and state reshuffling near avoided crossings.
These results demonstrate the sensitivity of hadron structure to external magnetic fields and help bridge our understanding to relativistic approaches.

\end{abstract}

\maketitle

%=========================================================
\section{Introduction}
%=========================================================

The behavior of quarkonia under extreme conditions---including strong magnetic fields, finite temperatures, and densities---provides crucial insights into the nonperturbative structure of Quantum Chromodynamics (QCD) vacuum. This field has attracted growing interest since the original proposal of $J/\psi$ suppression as a quark-gluon plasma signature~\cite{Matsui:1986dk,Hashimoto:1986nn}. The relevance has been further amplified by relativistic heavy-ion collisions, where non-central collisions generate transient but intense magnetic fields reaching $|eB| \sim 0.3~\text{GeV}^2$ at the LHC~\cite{Kharzeev:2007jp,Skokov:2009qp,Voronyuk:2011jd}, capable of significantly modifying hadronic properties.

Recent theoretical advances have enabled comprehensive studies of these phenomena through multiple approaches. Lattice QCD simulations now provide first-principles calculations of hadron properties in magnetic fields~\cite{Bali:2011qj,Luschevskaya:2012xd,Hidaka:2012mz,Luschevskaya:2014lga,Luschevskaya:2015cko,Bali:2017ian,Bali:2018sey,Luschevskaya:2018chr,Hattori:2019ijy,Bignell:2019vpy,Bignell:2020dze,Ding:2020hxw,Ding:2022tqn,Ding:2025pbu} (\cite{Endrodi:2024cqn} for a review), while model-based frameworks offer complementary perspectives. This vibrant research area is evidenced by numerous recent reviews covering various aspects of quarkonia in extreme environments~\cite{Kharzeev:2012ph,Mocsy:2013syh,Andersen:2014xxa,Huang:2015oca,Hattori:2016emy,Fukushima:2018grm,Rothkopf:2019ipj,Zhao:2020jqu,Iwasaki:2021nrz,Hattori:2023egw,Adhikari:2024bfa}.

The study of heavy hadrons containing charm or bottom quarks under magnetic fields has been approached through several theoretical frameworks. 
Constituent quark models~\cite{Machado:2013rta,Alford:2013jva,Bonati:2015dka,Suzuki:2016kcs,Yoshida:2016xgm,Iwasaki:2018pby,Wen:2025dwy} and QCD sum rules~\cite{Cho:2014exa,Cho:2014loa,Gubler:2015qok,Parui:2022msu} have provided valuable insights, complemented by holographic QCD approaches~\cite{Dudal:2014jfa,Braga:2018zlu,Braga:2019yeh,Braga:2020hhs,Jena:2022nzw} and the complex-potential picture in thermal medium~\cite{Hasan:2017fmf,Singh:2017nfa,Hasan:2020iwa,Ghosh:2022sxi,Sebastian:2023tlw}.
Lattice QCD simulations have further advanced our understanding by investigating the heavy quark-antiquark potential in magnetic backgrounds~\cite{Bonati:2014ksa,Bonati:2016kxj,DElia:2021tfb}.
Beyond heavy-quark systems, significant attention has been devoted to understanding magnetic field effects on hadrons composed of light (up, down, or strange) quarks.
For light mesons, studies (e.g., \cite{Chernodub:2010qx,Chernodub:2011mc,Callebaut:2011ab,Ammon:2011je,Chernodub:2011gs,Kojo:2012js,Andreichikov:2013zba,Taya:2014nha,Hattori:2015aki,Li:2016gfn,Coppola:2018vkw,Kojo:2021gvm}) have revealed characteristic modifications, while light baryons have been examined through several approaches~\cite{Andreichikov:2013pga,Haber:2014ula,Taya:2014nha,He:2015zca,He:2016oqk,He:2018vfc,Yakhshiev:2019gvb,Coppola:2020mon,Chen:2023gws,Chen:2023jbq,Samanta:2025mrq}.

The constituent quark model remains a powerful framework for understanding static hadron properties, as detailed in a recent review~\cite{Iwasaki:2021nrz}. Magnetic fields influence hadrons through three main mechanisms: (i) spin mixing via the $-\bm{\mu}_i \cdot \bm{B}$ interaction, (ii) quark kinetic energy modifications induced by the $\bm{B} \times \bm{r}$ term, and (iii) anisotropic deformation of inter-quark potentials~\cite{Miransky:2002rp,Andreichikov:2012xe,Chernodub:2014uua,Rougemont:2014efa}, where lattice QCD simulations provide crucial constraints~\cite{Bonati:2014ksa,Bonati:2016kxj,DElia:2021tfb}. 
Previous investigations of quarkonia~\cite{Suzuki:2016kcs,Yoshida:2016xgm} have systematically examined these effects, particularly spin mixing and energy shifts, employing potential models. These studies utilize the cylindrical Gaussian expansion method (CGEM) to solve the two-body Schr\"odinger equation, an adaptation of the conventional Gaussian expansion method (GEM)~\cite{Kamimura:1988zz,Hiyama:2003cu} specifically designed for systems with cylindrical symmetry under magnetic fields.

Building upon previous studies~\cite{Yoshida:2016xgm,Suzuki:2016kcs}, we aim to study light-front wave functions (LFWFs) of heavy quarkonia from nonrelativistic quark model solutions under strong magnetic fields and analyze their structural modifications. 
Unlike the direct light-front Hamiltonian diagonalization method proposed in Ref.~\cite{Wen:2025dwy}, our approach establishes a connection between magnetically-modified LFWFs and the rest-frame wave functions, which offers clearer physical interpretation in coordinate space. 
From these LFWFs, one can directly access partonic observables such as parton distribution functions (PDFs), relevant for high-energy processes~\cite{Lepage:1980fj,Brodsky:1997de,Diehl:2003ny}. 
Connections between rest-frame wave functions and LFWFs, particularly in the absence of a magnetic field, have been extensively utilized in previous studies (e.g.,~\cite{Cardarelli:1995dc,Choi:1997iq, Choi:2015ywa, Dhiman:2019ddr, Arifi:2022pal,Ridwan:2024ngc, Arifi:2024mff, Wu:2025rto}), providing a solid foundation for the current work.

The present study has several key findings:  
(i) These LFWFs exhibit common features also seen in relativistic models, 
such as transverse momentum broadening.
This behavior can be clearly understood from the squeezed wave function in transverse coordinate space~\cite{Suzuki:2016kcs,Yoshida:2016xgm}.
(ii) The ground-state PDFs, characterizing the longitudinal momentum fraction distributions, exhibit only minor modifications at leading order. 
Substantial modifications emerge at next-to-leading order through corrections.
In contrast, the PDF for excited states is strongly modified due to nodal structure deformation.
While experimental verification remains challenging, our predictions provide useful input for lattice QCD simulations~\cite{Endrodi:2024cqn}, especially with recent developments in Large Momentum Effective Theory (LaMET)~\cite{Ji:2013dva}, which enables direct access to partonic structure. 

This paper is organized as follows. Sec.~\ref{sec:method} introduces the two-body Hamiltonian, the CGEM framework, and the construction of LFWFs. Numerical results and their detailed analysis are presented in Sec.~\ref{sec:result}. The conclusions and future outlook are provided in Sec.~\ref{sec:conclusion}. Additional details concerning relativistic corrections and the explicit forms of matrix elements are given in Appendices~\ref{app:correction} and~\ref{app:matrix_element}, respectively.

%=========================================================
\section{Formalism} 
%=========================================================
\label{sec:method}

In this section, we construct the effective Hamiltonian
and introduce the CGEM to solve the Scr\"{o}dinger with the cylindrical symmetry constraint. 
The LFWF is then introduced and some related quantities are discussed.

%---------------------------------------------------------
\subsection{Quark model in a magnetic field}
%---------------------------------------------------------

In this study, we investigate quarkonia in the nonrelativistic quark model.  
In the presence of an external magnetic field, the two-body Hamiltonian composed of a quark and an antiquark can be expressed as~\cite{Machado:2013rta, Alford:2013jva}
%#######################################  
\begin{equation}
    H_{q\bar{q}} = \sum_{i=q,\bar{q}} \left[ m_i + \frac{(\bm{p}_i - q_i \bm{A})^2}{2m_i} - \bm{\mu}_i \cdot \bm{B} \right] + V_{q\bar{q}}(r).
\end{equation}
%#######################################  
Here, $m_i$ and $\bm{p}_i$ denote the mass and momentum of quark or antiquark labeled by $i=q,\bar{q}$, respectively.  
The magnetic field is contained in the Hamiltonian through the vector potential $\bm{A}$, which alters the kinetic-energy term of the single particle.  
We adopt the symmetric gauge, $\bm{A}(\bm{r}_i) = \frac{1}{2} \bm{B} \times \bm{r}_i$, and include the coupling to the quark magnetic moment $\bm{\mu}_i$.  

For the effective quark-antiquark potential $V_{q\bar{q}}$, we adopt the following form~\cite{Eichten:1974af}
%#######################################
\begin{eqnarray}\label{eq:potential}
V_{q\bar{q}} &=& C + \sigma r- \frac{A}{r} + \alpha (\bm{S}_q \cdot \bm{S}_{\bar{q}}) \mathrm{e}^{-\Lambda r^2},\quad
\end{eqnarray}
%#######################################
where $r=\sqrt{ z^2 + r_\perp^2}$, $C$ is a constant shift, and the magnitudes of the confining, 
color Coulomb, and spin-spin potentials are characterized by the parameters, $\sigma$, $A$, and $\alpha$, respectively.
We smear the spin-spin interaction as a Gaussian function with a parameter $\Lambda$~\cite{Barnes:2005pb}, for simplicity.
The term $\Braket{\bm{S}_q \cdot \bm{S}_{\bar{q}} }$ yields the values of $1/4$ and $-3/4$ 
for the vector (spin-triplet) and pseudoscalar (spin-singlet) mesons, respectively.

In a magnetic field, the kinetic momentum $\bm{p}_i$ is not conserved, while one can introduce the so-called pseudo-momentum~\cite{Johnson:1949}.
The two-body pseudo-momentum is defined as~\cite{Gor'kov:1968,Carter:1969aj,Avron:1978}
\begin{eqnarray}
    \bm{K} = \sum_{i=q,\bar{q}} \left[\bm{p}_i + \frac{1}{2}q_i \bm{B}\cross \bm{r}_i \right],
\end{eqnarray}
which is a conserved quantity. 
By using the pseudo-momentum, the total wave function can be factorized as 
\begin{eqnarray}
    \Phi(\bm{R},\bm{r}) &=& \exp\left[i\bm{P}\cdot \bm{R}\right]\psi(\bm{r})  \\
    &=& \exp\left[i(\bm{K} - \frac{q}{2} \bm{B}\cross \bm{r})\cdot \bm{R}\right]\psi(\bm{r}),
\end{eqnarray}
where $\bm{R}$ and $\bm{r}$ are the center-of-mass and relative coordinate, respectively.
$\bm{P}$ is the center-of-mass momentum.

Here we focus on the magnetic field $B$ along the $z$ direction and meson with $\bm{K}=0$.
We also focus on heavy quarkonia with neutral charge, $q_1 = -q_2 = q$. 
In this case, the two-body Hamiltonian is reduced to the Hamiltonian for the relative motion~\cite{Alford:2013jva,Andreichikov:2013zba,Bonati:2015dka,Suzuki:2016kcs,Yoshida:2016xgm}
\begin{eqnarray}\label{eq:Hamiltonian}
    H = -\frac{\bm{\nabla}^2}{2\mu} + \frac{q^2B^2}{8\mu}r_\perp^2 - \sum_{i=q,\bar{q}} \bm{\mu}_i \cdot \bm{B} + V(r). \quad 
\end{eqnarray}
where $\mu= m_q/2$ is the reduced quark mass. 
The above Hamiltonian maintains the rotational symmetry on the transverse $x$-$y$ plane and the reflection symmetry along the $z$-axis.
The second term proportional to $B^2r_\perp^2$ leads to the so-called quark Landau levels.

We discuss the contribution of the effects induced by $\bm{\mu}_i \cdot \bm{B}$~\cite{Machado:2013rta,Alford:2013jva}, where $\bm{\mu}_i = {gq_i\bm{S}_i}/{2m_i}$ is the quark magnetic moment with the Lande $g$-factor, $g=2$.
Thus, we obtain 
\begin{eqnarray}
    -( \bm{\mu}_1 +  \bm{\mu}_2) \cdot \bm{B} = - \frac{gq}{4}\left(\frac{\bm{\sigma}_1}{m_1}-\frac{\bm{\sigma}_2}{m_2}\right) \cdot \bm{B}
\end{eqnarray}
where we used $\bm{S}_i=\bm{\sigma}_i/2$. 
Then, the eigenstates with the different spin quantum numbers, mix with each other 
\begin{eqnarray}
    \bra{10} -( \bm{\mu}_1 +  \bm{\mu}_2) \cdot \bm{B} \ket{00} &=& -\frac{gqB}{4\mu},\\
    \bra{00} -( \bm{\mu}_1 +  \bm{\mu}_2) \cdot \bm{B} \ket{10} &=& -\frac{gqB}{4\mu},
\end{eqnarray}
where the spin eigenstates are labeled by $\ket{SS_z}$ with $S=0,1$ and $S_z=0,\pm1$.
Thus, only the longitudinal component $\ket{10}$ of $S=1$ can mix with the $S=0$ state whereas the transverse components $\ket{1\pm1}$ do not mix with other components. 
Taking into account the off-diagonal interaction, we should solve a coupled-channel Schr\"{o}dinger equation. 
We note that in the case of quarkonia, all the diagonal matrix elements, such as $\bra{00}  -( \bm{\mu}_1 + \bm{\mu}_2) \cdot \bm{B} \ket{00}$, are vanishing: there is no hadronic Zeeman shift.

%.........................................................
\subsection{Cylindrical Gaussian Expansion Method}
%.........................................................

To diagonalize the Hamiltonian~\eqref{eq:Hamiltonian}, we use a cylindrical Gaussian expansion method~\cite{Suzuki:2016kcs,Yoshida:2016xgm}, which is a generalization of the Gaussian expansion method~\cite{Kamimura:1988zz,Hiyama:2003cu}.
In this method, we expand the wave function in terms of a set of Gaussian basis functions, $\phi^{CG}_n$, with a different Gaussian range parameter as
%#######################################
\begin{eqnarray} \label{eq:GEM}
     \psi &=& \sum_{n=1}^{n_\mathrm{max}} c_{n}   \phi_{n}^{\mathrm{CG}}, 
\end{eqnarray}
%#######################################
where $c_n$ represents the expansion coefficient.
The Gaussian basis function in position space is given by~\cite{Suzuki:2016kcs,Yoshida:2016xgm}
%#######################################
\begin{eqnarray}
    \phi^{\mathrm{CG}}_n(z,\bm{r}_\perp)  = N_n \mathrm{e}^{-\beta_n{r}_\perp^2}\mathrm{e}^{-\gamma_n z^2},
\end{eqnarray}
%#######################################
and the basis function in the momentum space, obtained through the Fourier transformation
\begin{eqnarray}
    \phi_{n}^{\mathrm{CG}} (p_z,\bm{p}_\perp) &=& \frac{1}{(2\pi)^{3/2}}\int_0^\infty \dd r_\perp r_\perp \int_{-\infty}^\infty \dd z \int_0^{2\pi}\dd\phi\nonumber\\
            & & \times \mathrm{e}^{-ip_z z}\mathrm{e}^{-i \bm{r}_\perp\cdot \bm{p}_\perp} \phi^{\mathrm{CG}}_n(z,\bm{r}_\perp),
\end{eqnarray}
is given by
%#######################################
\begin{eqnarray}
\phi_{n}^{\mathrm{CG}} (p_z,\bm{p}_\perp) &=& \tilde{N}_n \mathrm{e}^{-{p}_\perp^2/ 4\beta_n} \mathrm{e}^{-{p}_z^2/ 4\gamma_n} .
\end{eqnarray}
%#######################################
The normalization factors of the basis function, $N_n$ and $\tilde{N}_n$, are determined by $\braket{\phi_{n}^{\mathrm{CG}}|\phi_{n}^{\mathrm{CG}}}=1$.

In the GEM, two sets of parameters must be determined: the expansion coefficients $c_n$ and the Gaussian range parameters $(\beta_n, \gamma_n)$.
The coefficients $c_n$ are obtained by solving the generalized eigenvalue problem,  
\begin{equation}
    \bm{H}_{q\bar{q}}\, \bm{c} = M_{q\bar{q}}\, \bm{S}\, \bm{c},
\end{equation}  
where the Hamiltonian matrix elements are given by $H_{q\bar{q},nm} = \bra{\phi_n^\mathrm{CG}} \hat{H} \ket{\phi_m^\mathrm{CG}}$, and the elements of the overlap matrix are $S_{nm} = \braket{\phi_n^\mathrm{CG} | \phi_m^\mathrm{CG}}$.  
The Gaussian range parameters $\beta_n$ and $\gamma_n$ are determined by adopting a geometric progression~\cite{Kamimura:1988zz,Hiyama:2003cu}
%#######################################
\begin{align}
    \beta_n &= \frac{1}{r_{\perp n}^2}, r_{\perp n} = r_{\perp 1} b^{n-1},  b = \left(\frac{r_{\perp \mathrm{max}}}{r_{\perp1}}\right)^{\frac{1}{n_\mathrm{max}-1}},\\
    \gamma_n & = \frac{1}{z_n^2}, z_n = z_1 c^{n-1},  c = \left(\frac{z_\mathrm{max}}{z_1}\right)^{\frac{1}{n_\mathrm{max}-1}},
\end{align}
%#######################################
which reduces them to only four parameters ($\beta_1$, $\beta_{n_\mathrm{max}}$) and ($\gamma_1$, $\gamma_{n_\mathrm{max}}$) to be optimized.
We note that the basis functions are non-orthogonal $S_{nm}\neq \delta_{nm}$ and the states are normalized as
%#######################################
\begin{eqnarray}
  \braket{\psi | \psi} = \sum_{n,m} c_n^* S_{nm}c_m = 1.
\end{eqnarray}
%#######################################

Since the spin singlet $\ket{00}$ and longitudinal component $\ket{10}$ of 
the spin triplet mix with each other in the magnetic field, 
the wave function should take into account the coupled channel.
The basis function is given as a linear combination of spin-0 and spin-1 components~\cite{Suzuki:2016kcs,Yoshida:2016xgm} 
\begin{eqnarray}
    \Psi(r_\perp,z) = \sum_{n=1}^N c_n^{S=0}\Phi_n^{S=0} + \sum_{n=N+1}^{2N} c_n^{S=1}\Phi_n^{S=1}\quad
\end{eqnarray}
where 
\begin{eqnarray}
    \Phi_n^{S=0}(r_\perp,z) &=& N_n^P \mathrm{e}^{-\beta_n^P r_\perp^2} \mathrm{e}^{-\gamma_n^P z^2}, \\
    \Phi_n^{S=1}(r_\perp,z) &=& N_n^V \mathrm{e}^{-\beta_n^V r_\perp^2} \mathrm{e}^{-\gamma_n^V z^2}.
\end{eqnarray}
In this case, we have four range parameters 
\begin{eqnarray}
    (\beta_n^P,\gamma_n^P), \qquad (\beta_n^V,\gamma_n^V).
\end{eqnarray}
Because of the geometric progression applied, in total we have eight parameters to optimize.

%.........................................................
\subsection{Light-front wave functions}
%.........................................................

After diagonalizing the Hamiltonian~\eqref{eq:Hamiltonian}, the resulting wave functions are transformed into LFWFs~\cite{Cardarelli:1995dc, Choi:1997iq,Choi:2015ywa, Dhiman:2019ddr,Arifi:2022pal,Ridwan:2024ngc,Arifi:2024mff,Wu:2025rto}. 
These LFWFs are expressed in terms of internal variables that are invariant under Lorentz boosts in the longitudinal direction, defined as
\begin{align}
   x_i = \frac{p_i^+}{P^+},\qquad
   \bm{k}_{\perp i} = \bm{p}_{\perp i} - x_i \bm{P}_\perp,
\end{align}
where $P^\mu = (P^+, P^-, \bm{P}_\perp)$ is the meson's total four-momentum and $p_i^\mu$ denotes the four-momentum of the $i$-th constituent quark. Here, we denote on the quark’s longitudinal momentum fraction $x \equiv x_q$ and transverse momentum $\bm{k}_\perp \equiv \bm{k}_{\perp q}$.

To obtain the LFWFs, we first perform a change of variables from the momentum-space coordinates $(k_z, \bm{k}_\perp)$ to $(x, \bm{k}_\perp)$. 
The longitudinal momentum component $k_z$ can be expressed as
\begin{equation}\label{eq:k_z}
    k_z = \left(x - \frac{1}{2}\right) M_0 + \frac{m_{\bar{q}}^2 - m_q^2}{2 M_0},
\end{equation}
where the invariant meson mass \(M_0\) is given by
\begin{equation}
    M_0^2 = \frac{\bm{k}_\perp^2 + m_q^2}{x} + \frac{\bm{k}_\perp^2 + m_{\bar{q}}^2}{1 - x}.
\end{equation}
The radial part of the LFWF is defined as
\begin{equation}
    \Phi(x, \bm{k}_\perp) = \sqrt{2 (2\pi)^3} \, \sqrt{\frac{\partial k_z}{\partial x}} \, \psi(\bm{k}),
\end{equation}
where \(\psi(\bm{k})\) is the wave function obtained from the Hamiltonian diagonalization. The Jacobian factor associated with the variable transformation is
\begin{equation}
    \frac{\partial k_z}{\partial x} = \frac{M_0}{4 x (1 - x)} \left[ 1 - \frac{(m_q^2 - m_{\bar{q}}^2)^2}{M_0^4} \right].
\end{equation}

The spin and orbital angular momentum part $\mathcal{R}^{JJ_z}_{\lambda_q \lambda_{\bar{q}}}$ of the meson LFWF is obtained~\cite{Terentev:1976jk,Berestetsky:1976um,Berestetsky:1977zk} through the interaction-independent Melosh transformation~\cite{Melosh:1974cu}, which relates the instant-form spin states to the light-front helicity basis. 
Here $J$ and $J_z$ denote the total spin and its projection along the quantization axis, respectively, while $\lambda_q$ and $\lambda_{\bar{q}}$ are the light-front helicities of the quark and antiquark.
This procedure assumes that the meson is described in terms of a non-interacting quark–antiquark basis, 
in line with the Bakamjian–Thomas construction~\cite{Bakamjian:1953kh,Arifi:2022qnd,Arifi:2023uqc,Arifi:2025olq}. 
This procedure connects the spin-orbital structure of the ordinary rest-frame wave function, which carries definite quantum numbers $J^{PC}$, to its light-front representation in a consistent way. 
The effect of the external magnetic field is first included in the rest-frame wave function, by diagonalizing the Hamiltonian~\eqref{eq:Hamiltonian}. 
We assume that these modifications are then carried over to the LFWFs via the Melosh transformation.
Since our goal is to understand how the LFWFs are modified in the nonrelativistic quark model, this approach captures the dominant effects of the magnetic field such as spin-mixing and quark Landau levels.  

For the spin-orbital angular momentum part, it is convenient to work with the covariant form as~\cite{Jaus:1989au, Jaus:1991cy}
%#######################################
\begin{equation}
    \mathcal{R}^{JJ_z}_{\lambda_q \lambda_{\bar{q}}} = \frac{1}{\sqrt{2} \tilde{M}_0} 
    \bar{u}_{\lambda_q}(p_q)\, \Gamma_{\rm M}\, v_{\lambda_{\bar{q}}}(p_{\bar{q}}),
\end{equation}
%#######################################
where $\tilde{M}_0 \equiv \sqrt{M_0^2 - (m_q - m_{\bar{q}})^2}$, and $u(p_q)$ and $v(p_{\bar{q}})$ are the light-front Dirac spinors for the quark and antiquark, respectively, with the effect of the Melosh tranformation implicitly included through the spinor structure~\cite{Ji:1992yf}.
The vertex function $\Gamma_{\rm M}$ depends on the meson type, with
\begin{align}
\Gamma_{\rm P} = \gamma_5, \qquad \Gamma_{\rm V} = -\slashed{\epsilon}(J_z) + \frac{\epsilon \cdot (p_q - p_{\bar{q}})}{M_0 + m_q + m_{\bar{q}}},
\end{align}
where $\Gamma_{\rm P}$ and $\Gamma_{\rm V}$ correspond to pseudoscalar and vector mesons, respectively.
The polarization vectors $\epsilon^\mu(J_z) = (\epsilon^+, \epsilon^-, \bm{\epsilon}_\perp)$ are defined by
\begin{align}
\epsilon^\mu(\pm 1) &= \left(0, \frac{2 \bm{\epsilon}_\perp(\pm) \cdot \bm{P}_\perp}{P^+}, \bm{\epsilon}_\perp(\pm)\right), \\
\epsilon^\mu(0) &= \left(\frac{P^+}{M_0}, \frac{-M_0^2 + \bm{P}_\perp^2}{M_0 P^+}, \frac{\bm{P}_\perp}{M_0}\right),
\end{align}
with the transverse polarization vectors $\bm{\epsilon}_\perp(\pm 1) = (1, \pm i)/\sqrt{2}$.
Lastly, the spin-orbit wave function $\mathcal{R}^{JJ_z}_{\lambda_q \lambda{\bar{q}}}$ satisfies the normalization condition
\begin{equation}
\sum_{\lambda_q, \lambda_{\bar{q}}} \mathcal{R}^{JJ_z \dagger}_{\lambda_q \lambda_{\bar{q}}} \mathcal{R}^{J' J_z'}_{\lambda_q \lambda_{\bar{q}}} = \delta_{J J'} \delta_{J_z J_z'}.
\end{equation}
The LFWF of heavy quarkonia in momentum space can be expressed as
\begin{equation} \label{eq:LFWF}
    \varPsi^{JJ_z}_{\lambda_q \lambda_{\bar{q}}}(x, \bm{k}_\perp) = \Phi(x, \bm{k}_\perp) \, \mathcal{R}^{JJ_z}_{\lambda_q \lambda_{\bar{q}}}(x, \bm{k}_\perp),
\end{equation}
where \(\Phi(x, \bm{k}_\perp)\) represents the radial component of the wave function, and \(\mathcal{R}^{JJ_z}_{\lambda_q \lambda_{\bar{q}}}(x, \bm{k}_\perp)\) accounts for the spin and orbital angular momentum structure.
The LFWF density is defined by summing over the quark and antiquark helicities,~\cite{Wen:2025dwy}
\begin{equation} \label{eq:density}
    \rho(x, \bm{k}_\perp) = \frac{1}{2 (2\pi)^3} \sum_{\lambda_q, \lambda_{\bar{q}}} \left| \varPsi^{JJ_z}_{\lambda_q \lambda_{\bar{q}}}(x, \bm{k}_\perp) \right|^2.
\end{equation}
The wave function normalization condition is given by
\begin{equation}
    \int \mathrm{d}x \, \mathrm{d}^2 \bm{k}_\perp \, \rho(x, \bm{k}_\perp) = 1.
\end{equation}
We can also project the LFWF density to its longitudinal and transverse components, defined as 
\begin{eqnarray}\label{eq:pdf}
    \int \dd x f(x) = \int \dd k_\perp f(k_\perp) =1.
\end{eqnarray} 
It is worth noting that, in the vacuum, the longitudinal distribution $f(x)$ corresponds to the PDFs and the $\rho(x,\bm{k}_\perp)$ corresponds to the transverse momentum dependent distribution (TMD)~\cite{Lorce:2014hxa,Choi:2024ptc}.

%---------------------------------------------------------

\begin{figure*}[t]
    \centering
    \includegraphics[width=0.44\textwidth]{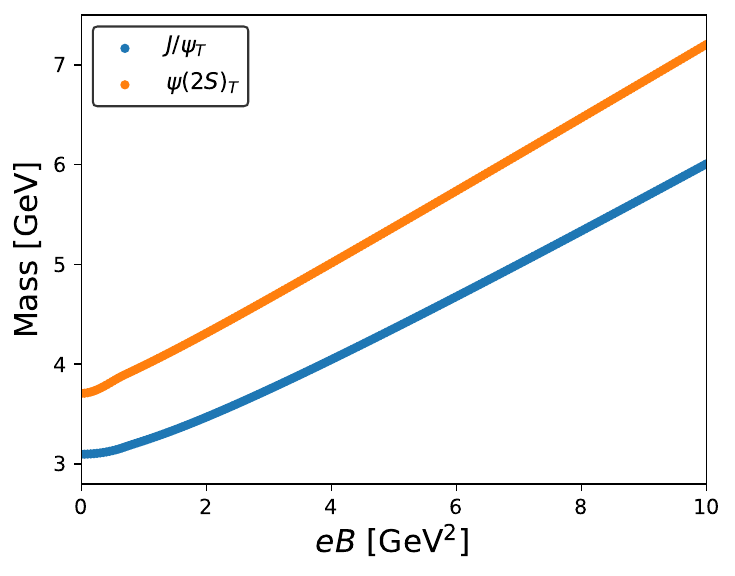}
    \includegraphics[width=0.45\textwidth]{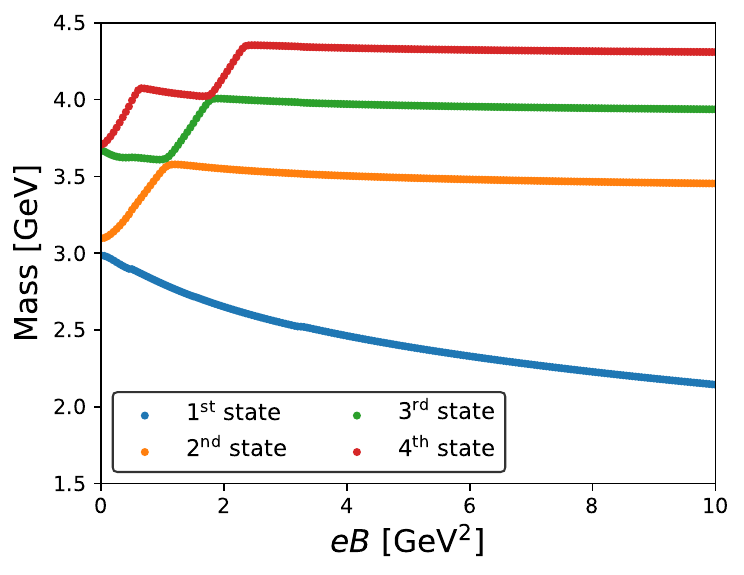}
    \caption{Magnetic-field dependence of the mass spectra: (left panel) $J/\psi_T$ and $\psi(2S)_T$ states; (right panel) $\eta_c(1S, 2S)$, $J/\psi_L$, and $\psi(2S)_L$ states. The Landau level primarily affects the masses of the transverse vector charmonia. Level repulsion and avoided crossings appear in the right panel due to state mixing. 
    Note that we plot a dense set of data points so that the curves appear continuous. }
    \label{fig:charm_mass}
\end{figure*}

%---------------------------------------------------------

As we will see later that the LFWFs become deformed under strong magnetic fields, it is useful to introduce the anisotropy parameter defined by
\begin{equation} \label{eq:anisotropy}
    \epsilon_{\mathrm{LF}} = \frac{\langle k_\perp^2 \rangle}{2 \langle k_z^2 \rangle} - 1,
\end{equation}
where $\langle k_\perp^2 \rangle = \langle k_x^2 + k_y^2 \rangle$. The factor of 2 in the denominator ensures a consistent comparison between the two-dimensional transverse and one-dimensional longitudinal momentum components. 
The longitudinal momentum operator $k_z$ is defined in Eq.~\eqref{eq:k_z}, and the expectation values $\langle k_\perp^2 \rangle$ and $\langle k_z^2 \rangle$ are determined by the Gaussian range parameters $\gamma$ and $\beta$ of the wave function.
For a spherical Gaussian basis, the relation $\langle k_\perp^2 \rangle = 2 \langle k_z^2 \rangle$ holds exactly, resulting in $\epsilon_{\mathrm{LF}} = 0$ and reflecting spherical symmetry. However, this relation no longer holds when using a cylindrical Gaussian basis, indicating anisotropic deformation of the LFWFs.

%=========================================================
\subsection{Numerical Setup} 
%=========================================================

For the parameters of the constituent quark model, in this work, we adopt those used in Ref.~\cite{Yoshida:2016xgm}, where some parameters were determined from lattice QCD simulations of charmonia~\cite{Kawanai:2011jt, Kawanai:2015tga}, as listed in Table~\ref{tab:parameter}. 
These parameters well reproduce the experimental masses~\cite{ParticleDataGroup:2024cfk} of heavy quarkonia in vacuum, as shown in Table~\ref{tab:vacuum_mass}.

We note that the model parameters may depend on the magnetic field.
For example, we can expect anisotropies in the confinement and Coulomb potentials (see Refs.~\cite{Miransky:2002rp,Andreichikov:2012xe,Chernodub:2014uua,Chernodub:2014uua,Rougemont:2014efa} for early works).
Such effects can be implemented in the potential model~\cite{Bonati:2015dka}.
However, since in the current work our aim is to examine the effects of the magnetic field on the LFWFs, we focus on the parameters without such modification. 

%%%%%%%%%%%%%%%%%%%%%%%%%%%%%%%%%%%%%%%%%%%%%%%%%%%%%%%%%%
\begin{table}[h] 
\begin{ruledtabular}
    \caption{ Parameters of constituent quark model~\cite{Yoshida:2016xgm}.} 
    \label{tab:parameter}
    \begin{tabular}{c|cccccc}
                & $m_Q$   & $\alpha$ [GeV] & $\sqrt{\sigma}$ [GeV] & $\Lambda$ [GeV$^2$] & $A$ & $C$ [GeV]\\ \hline 
    $(c\bar{c})$ & 1.784  & 0.4778 & 0.402 & 1.020 & 0.713 & $-0.5693$ \\
    $(b\bar{b})$ & 4.808  & 0.1322 & 0.402 & 1.020 & 0.531 & 0 \\
      \end{tabular}
\end{ruledtabular}
\end{table}
%%%%%%%%%%%%%%%%%%%%%%%%%%%%%%%%%%%%%%%%%%%%%%%%%%%%%%%%%%

%%%%%%%%%%%%%%%%%%%%%%%%%%%%%%%%%%%%%%%%%%%%%%%%%%%%%%%%%%
\begin{table}[h] 
\begin{ruledtabular}
    \caption{ Comparison between the predicted masses~\cite{Yoshida:2016xgm} and corresponding experimental data~\cite{ParticleDataGroup:2024cfk}. All the units are in GeV.} 
    \label{tab:vacuum_mass}
    \begin{tabular}{c|ccc|ccc}
                 & Meson          &  Our  & Expt. & Meson           &  Our  & Expt. \\ \hline 
    $(c\bar{c})$ & $\eta_c(1S)$   & 2.984 & 2.984 & $J/\psi$        & 3.097 & 3.097 \\
                 & $\eta_c(2S)$   & 3.669 & 3.638 & $\psi(2S)$      & 3.707 & 3.686 \\
    $(b\bar{b})$ & $\eta_b(1S)$   & 9.398 & 9.398 & $\Upsilon(1S)$  & 9.460 & 9.460 \\
                 & $\eta_b(2S)$   & 9.999 & 9.999 & $\Upsilon(2S)$  & 10.013 & 10.023 \\
                 & $\eta_b(3S)$   & 10.330 & \dots & $\Upsilon(3S)$  & 10.339 & 10.355 \\
      \end{tabular}
\end{ruledtabular}
\end{table}
%%%%%%%%%%%%%%%%%%%%%%%%%%%%%%%%%%%%%%%%%%%%%%%%%%%%%%%%%%

%=========================================================
\section{Results and Discussion} 
%=========================================================
\label{sec:result}

In this section, we investigate how the structures of charmonia and bottomonia are modified in the presence of strong magnetic fields. 
For each quarkonium, we first present the magnetic-field-induced mass shifts, recalculated using a denser set of data points than in previous studies~\cite{Yoshida:2016xgm}.
In this work, we specifically focus on examining the corresponding LFWF densities.
We also discuss how the transverse and longitudinal momentum distributions change and quantify the deformation using momentum expectation values and an anisotropy parameter. 
Finally, we discuss a possible relativistic correction, which plays an important role in modifying the longitudinal structure of the LFWFs, found using the light-front Hamiltonian approach~\cite{Wen:2025dwy}.

%---------------------------------------------------------
\subsection{Charmonia}
%---------------------------------------------------------

The mass spectra of charmonia under strong magnetic fields are shown in Fig.~\ref{fig:charm_mass}. 
These results were previously discussed in Ref.~\cite{Yoshida:2016xgm}. 
Here, we recalculate the magnetic field dependence of the mass spectra using a denser set of data points. 
One crucial aspect is the optimization procedure for the range parameters, $[\beta_1,\beta_{n_\mathrm{max}}]$ and $[\gamma_1,\gamma_{n_\mathrm{max}}]$. 
Since the wave function becomes deformed into a cigar-like shape under strong magnetic fields, the optimal range parameters vary with $eB$.
With improved optimization by Optim.jl~\cite{mogensen2018optim} for each data point, we confirm the magnetic field dependence of the mass spectra reported in Ref.~\cite{Yoshida:2016xgm}.
Although the nonrelativistic approximation works well when the magnetic field is smaller than the quark mass, we plot the mass spectra up to $eB=10$ GeV$^2$ for theoretical exploration, where the difference from the relativistic approach can be useful to discuss the relativistic effect.

The masses of the transverse vector charmonia ($J/\psi_T$ and $\psi(2S)_T$) are displayed in the left panel of Fig.~\ref{fig:charm_mass}, showing that they are mainly affected by the quark Landau levels. 
As a result, their mass shifts increase approximately linearly with the magnetic field. 
The mass shift for the excited state is larger than for the ground state, 
because the elongation of the spatial wave function in the $z$ direction leads to a larger expectation value of $\expval{r_\perp^2}$.

%---------------------------------------------------------

\begin{figure*}[t]
    \centering
    \includegraphics[width=0.92\textwidth]{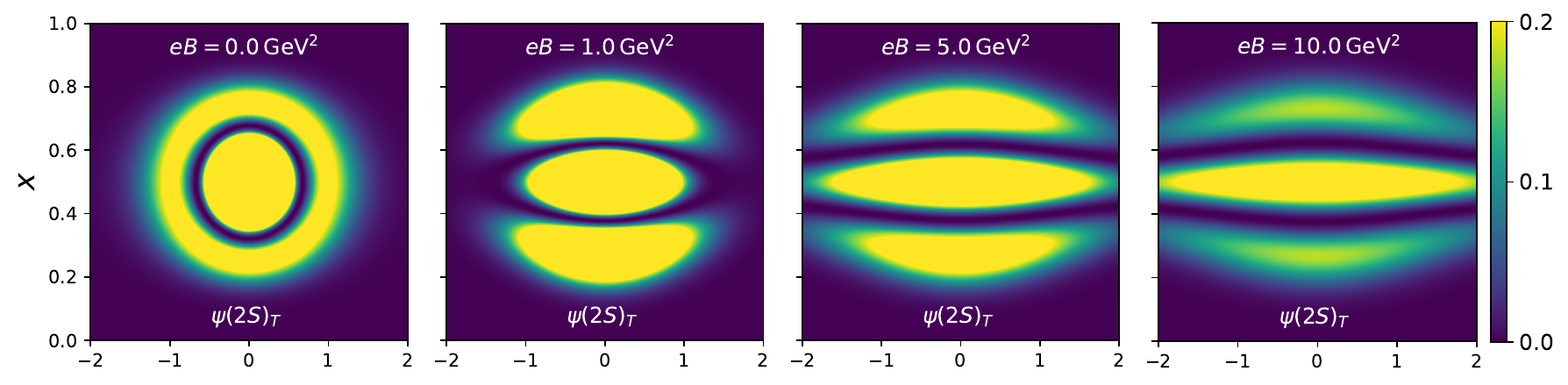}
    \includegraphics[width=0.92\textwidth]{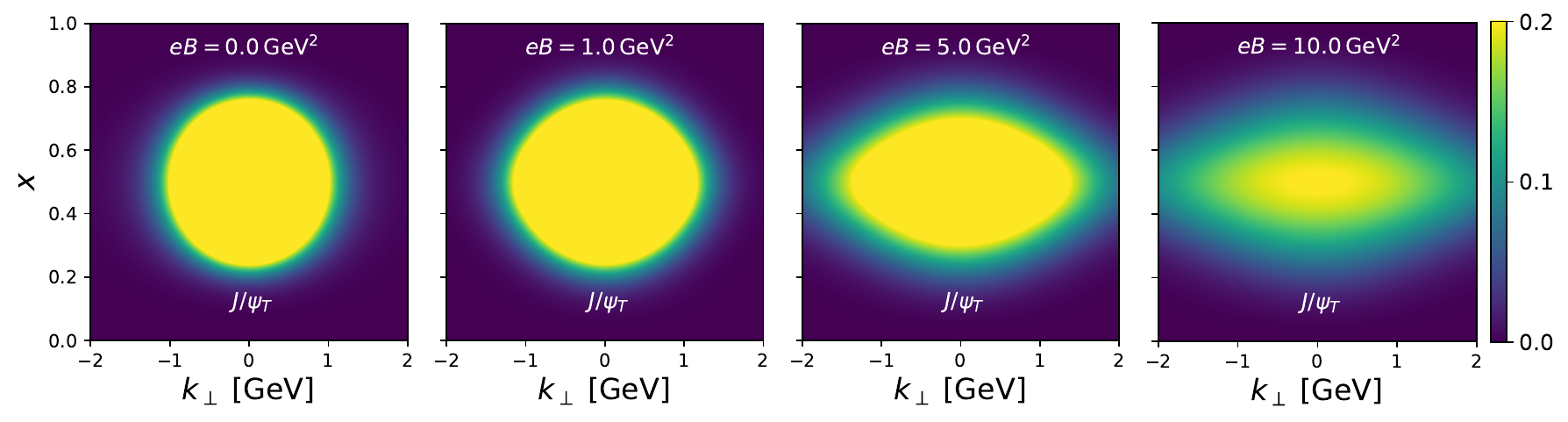}
    \caption{LFWF densities of the transverse vector charmonia at several different magnetic field strengths: (Top panels) $J/\psi$ states and (Bottom panels) $\psi(2S)$ states. The densities become more elongated in the transverse direction as the magnetic field increases.}
    \label{fig:charm_LFWF_T}
\end{figure*}

\begin{figure*}[t]
    \centering
    \includegraphics[width=0.92\textwidth]{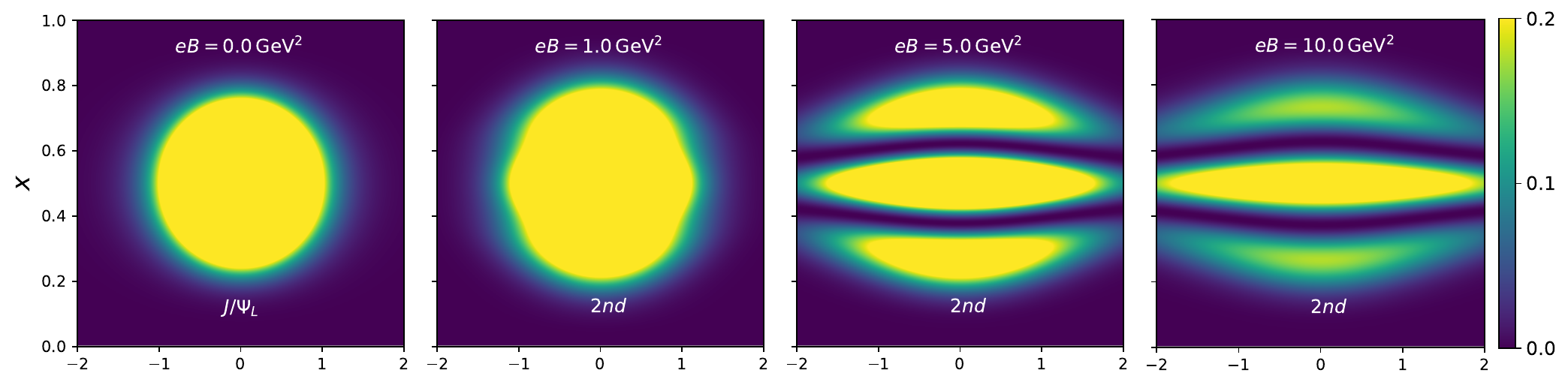}
    \includegraphics[width=0.92\textwidth]{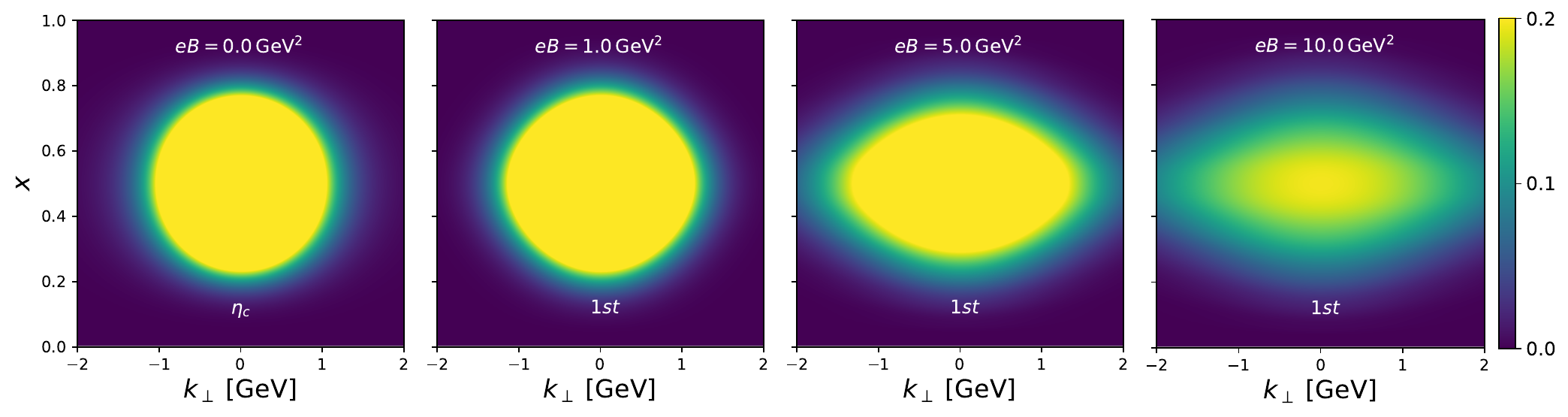}
    \caption{LFWF densities of pseudoscalar and longitudinal vector charmonia at several different magnetic field strengths: (Top panels) $2^\text{nd}$ state and (Bottom panels) $1^\text{st}$ state. The shape of the $2^\text{nd}$ state at certain magnetic fields changes significantly due to state mixing and avoided crossing.  }
    \label{fig:charm_LFWF_L}
\end{figure*}

\begin{figure*}[t]
    \centering
    \includegraphics[width=0.92\textwidth]{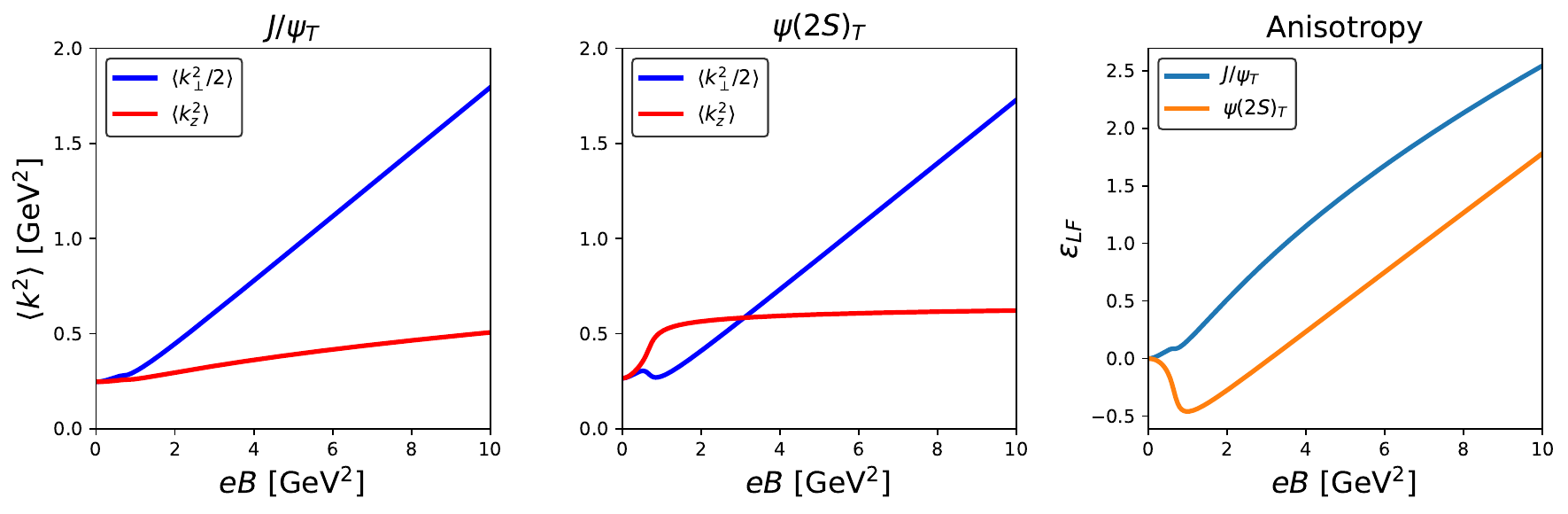}
    \caption{Magnetic-field dependence of $\expval{k^2}$ and the anisotropy parameter $\epsilon_{\mathrm{LF}}$ for $J/\psi_T$ and $\psi(2S)_T$ states. Transverse momentum broadening is one of the dominant effects of the magnetic field. The rapid increase in the longitudinal momentum of $\psi(2S)_T$ at low magnetic field arises from the redistribution of the LFWF density.
    }
    \label{fig:charm_asym}
\end{figure*}

\begin{figure}[b]
    \centering
    \includegraphics[width=0.46\textwidth]{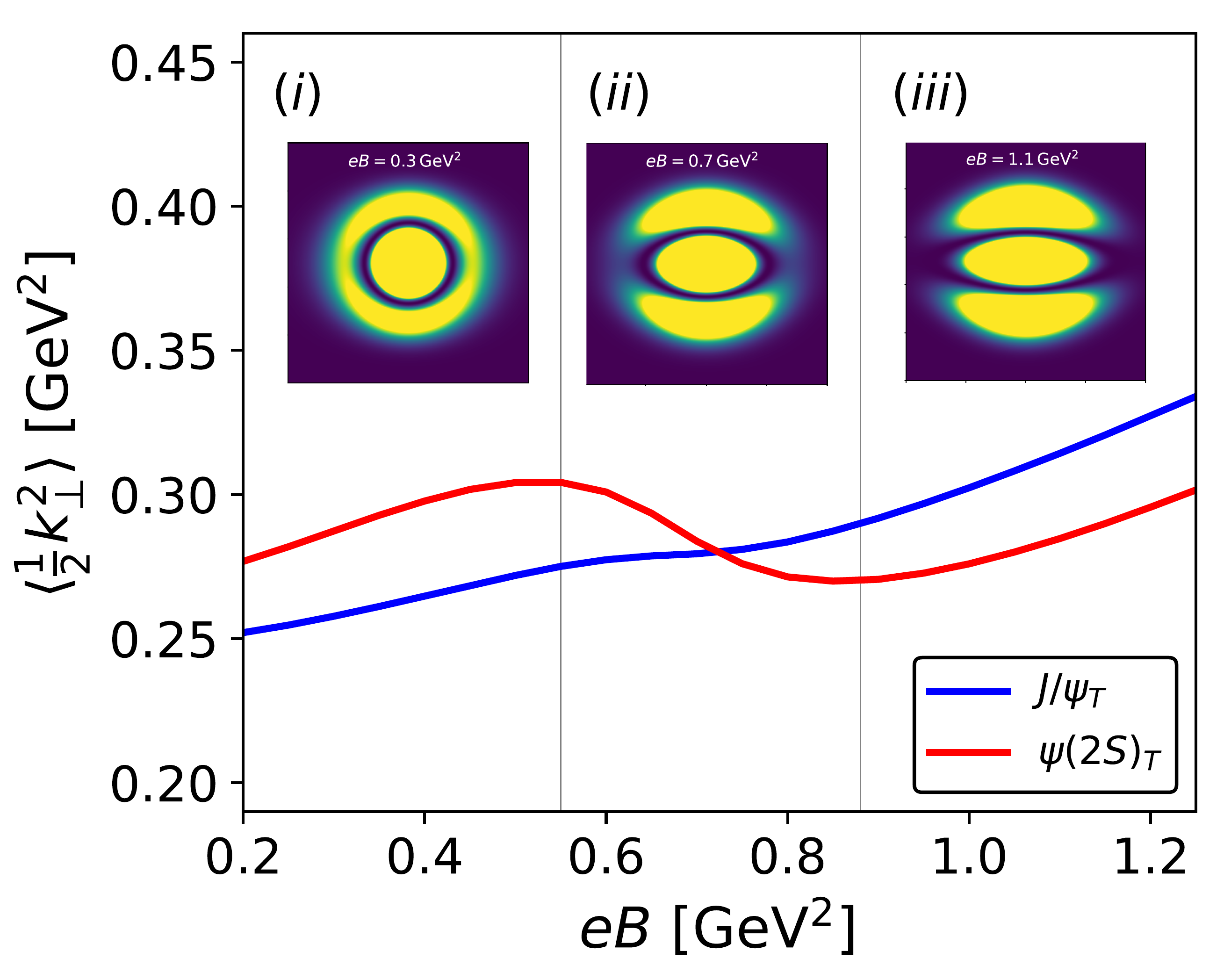}
    \caption{Close-up view of the weak-field region in Fig.~\ref{fig:charm_asym}, where $\left\langle \tfrac{1}{2}k_\perp^2 \right\rangle$ for $J/\psi_T$ and $\psi(2S)_T$ are plotted, and the typical LFWFs of $\psi(2S)_T$ in three different regimes: (i) momentum increase due to elongation of LFWFs {\it with transverse nodes}, (ii) momentum decrease due to the node disappearance, and (iii) momentum increase due to elongation of LFWFs {\it with no transverse node}.}
    \label{fig:charm_landau}
\end{figure}

\begin{figure*}[t]
    \centering
    \includegraphics[width=0.92\textwidth]{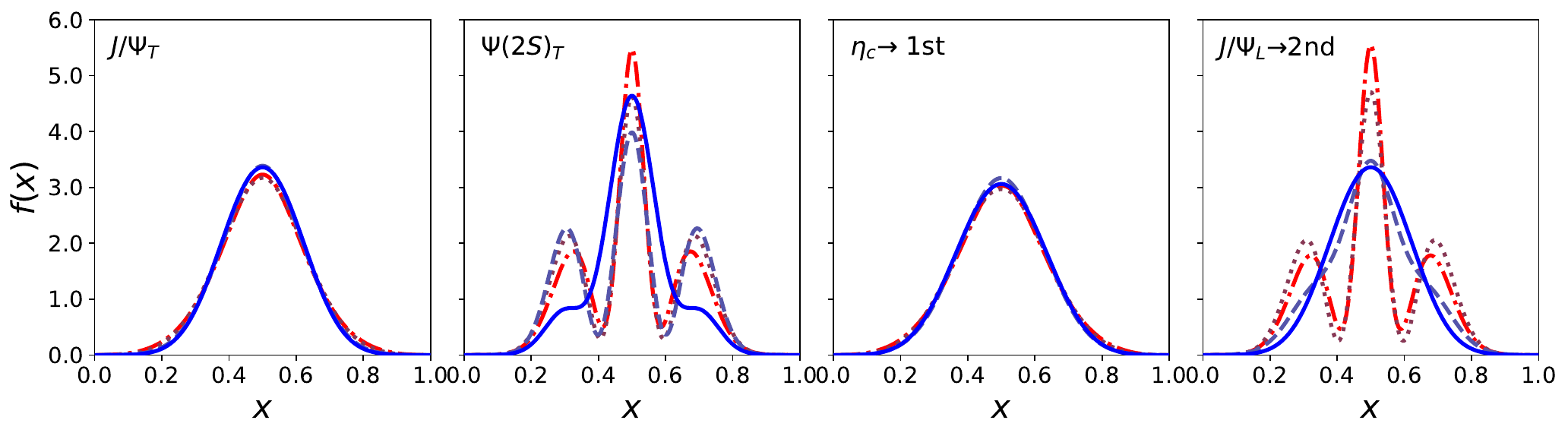}
    \includegraphics[width=0.92\textwidth]{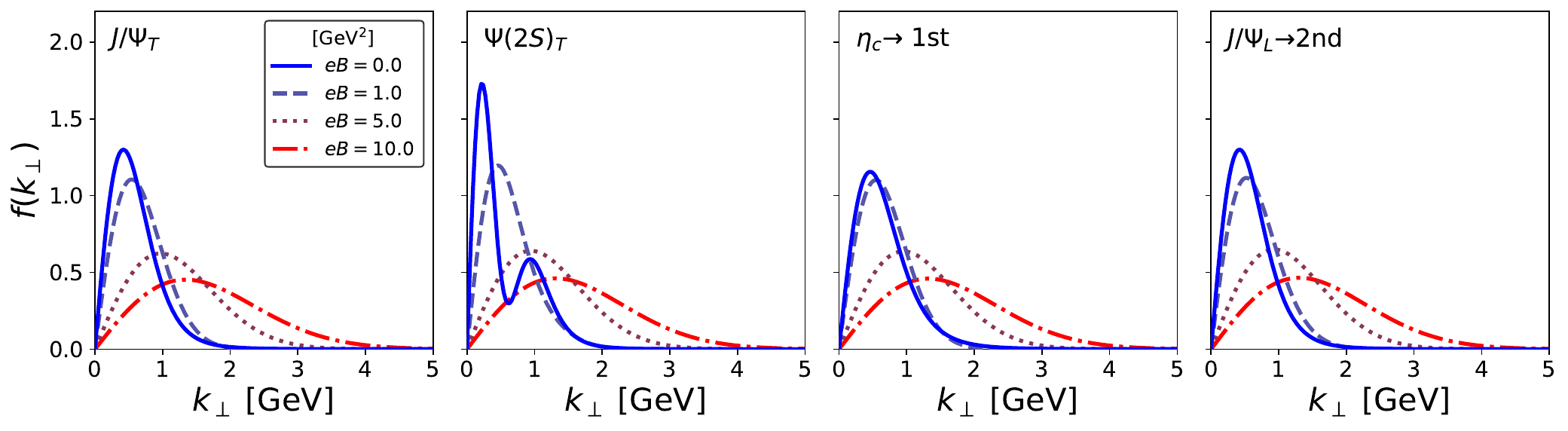}
    \caption{Longitudinal $f(x)$ and transverse $f(k_\perp)$ distributions of selected charmonia at several magnetic field strengths. The $f(x)$ is barely modified for the ground states while it is modified for the excited state due to the shape transition. The transition due to avoided crossing changes the distribution significantly. The $f(k_\perp)$ is generally broaden in the increasing magnetic field.}
    \label{fig:charm_pdf}
\end{figure*}

%---------------------------------------------------------

In contrast, the mass shifts of the pseudoscalar states ($\eta_c(1S)$ and $\eta_c(2S)$) and longitudinal vector states ($J/\psi_L$ and $\psi(2S)_L$) are shown in the right panel of Fig.~\ref{fig:charm_mass}. 
In this case, the pseudoscalar-vector mixing occurs due to the $-\bm{\mu}_i \cdot \bm{B}$ term, so that we label the resulting states as 1st, 2nd, 3rd, and 4th. 
The 1st and 2nd states decrease and increase, respectively, as a result of level repulsion with increasing magnetic field.
Around $eB = 1$–$3$ GeV$^2$ we find avoided crossings such as 2nd-3rd and 3rd-4th. We also note that in the strong magnetic field region, the level structure becomes simpler.

While the wave function density in coordinate space was discussed in Ref.~\cite{Yoshida:2016xgm}, here we focus on the LFWF density~\eqref{eq:density}. 
The LFWF densities for transverse charmonia at various $eB$ values are shown in Fig.~\ref{fig:charm_LFWF_T}. 
Since the LFWF is plotted in momentum space, the transverse momentum distribution becomes more extended with increasing $eB$, corresponding to the narrowing of the wave function in transverse coordinate space. 
This transverse momentum broadening originates from the magnetic-field–induced potential proportional to $B^2 r_\perp^2$ in Eq.~\eqref{eq:Hamiltonian}. 
Importantly, the nodal structure in the excited states at $eB=0$ is deformed toward the longitudinal direction. 
As a result, the density for excited states transforms into three well-separated oblate peaks.

The LFWF densities for the pseudoscalar and longitudinal vector charmonia are shown in Fig.~\ref{fig:charm_LFWF_L}, where we present several results for the 1st and 2nd states. 
The modifications of densities are similar to those in the transverse case, except that a more drastic change appears at certain values of $eB$. 
For example, the density of the 2nd state evolves from a single peak into three separated peaks.
Such a shape transition indicates that the energy of the radially excited state becomes lower than that of the nodeless state, as expected from the mass spectra in Fig.~\ref{fig:charm_mass}.

%---------------------------------------------------------

\begin{figure*}[t]
    \centering
    \includegraphics[width=0.47\textwidth]{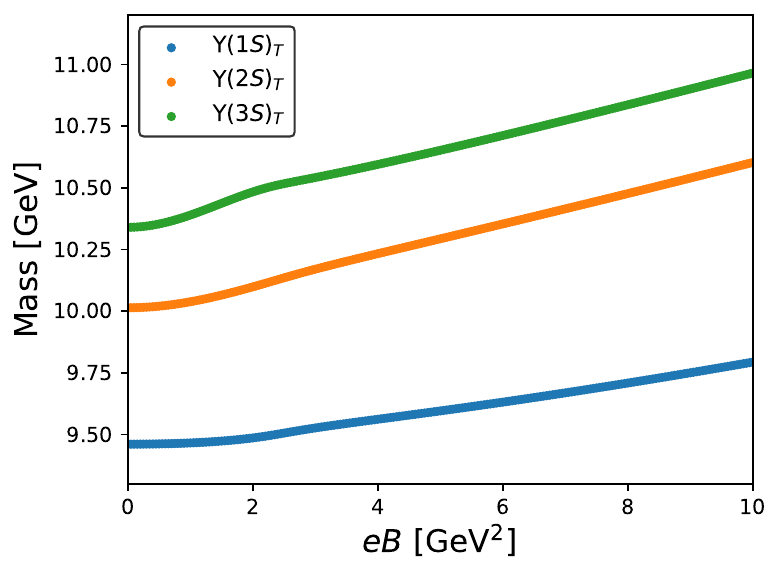}
    \includegraphics[width=0.47\textwidth]{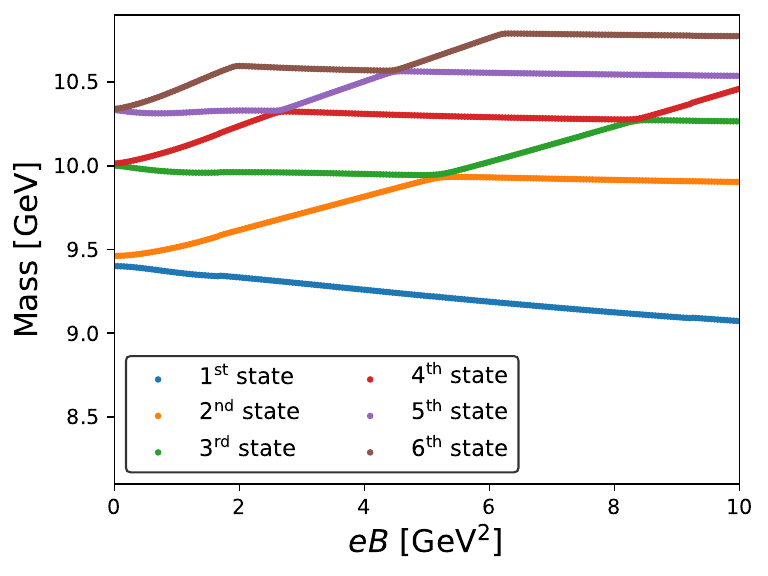}
    \caption{Magnetic-field dependence of the mass spectra: (left panel) $\Upsilon(1S, 2S, 3S)_T$ states; (right panel) $\eta_b(1S, 2S, 3S)$, $\Upsilon(1S, 2S, 3S)_L$ states. The Landau level primarily affects the masses of the transverse vector bottomonia. Level repulsion and avoided crossings appear in the right panel due to mixing between pseudoscalar and longitudinal vector states. Although the curves appear to overlap near the crossings, they remain distinct. }
    \label{fig:bottom_mass}
\end{figure*}

\begin{figure*}[t]
    \centering
    \includegraphics[width=0.95\textwidth]{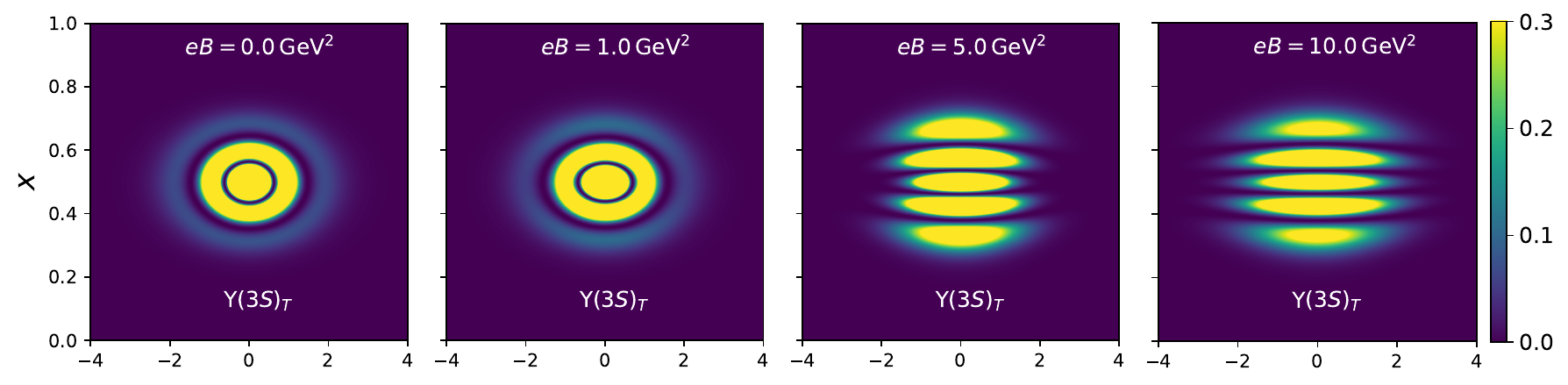}
    \includegraphics[width=0.95\textwidth]{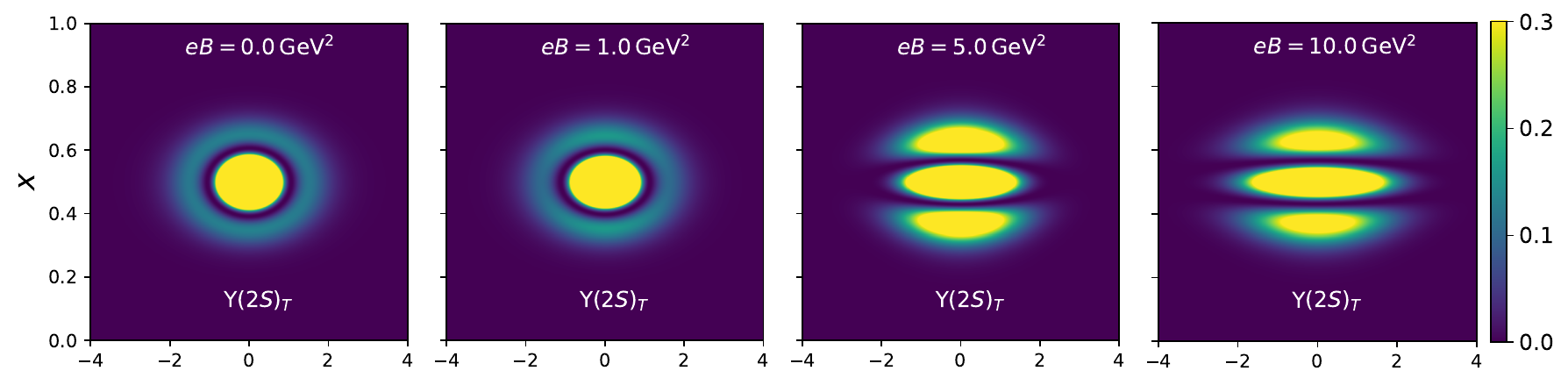}
    \includegraphics[width=0.95\textwidth]{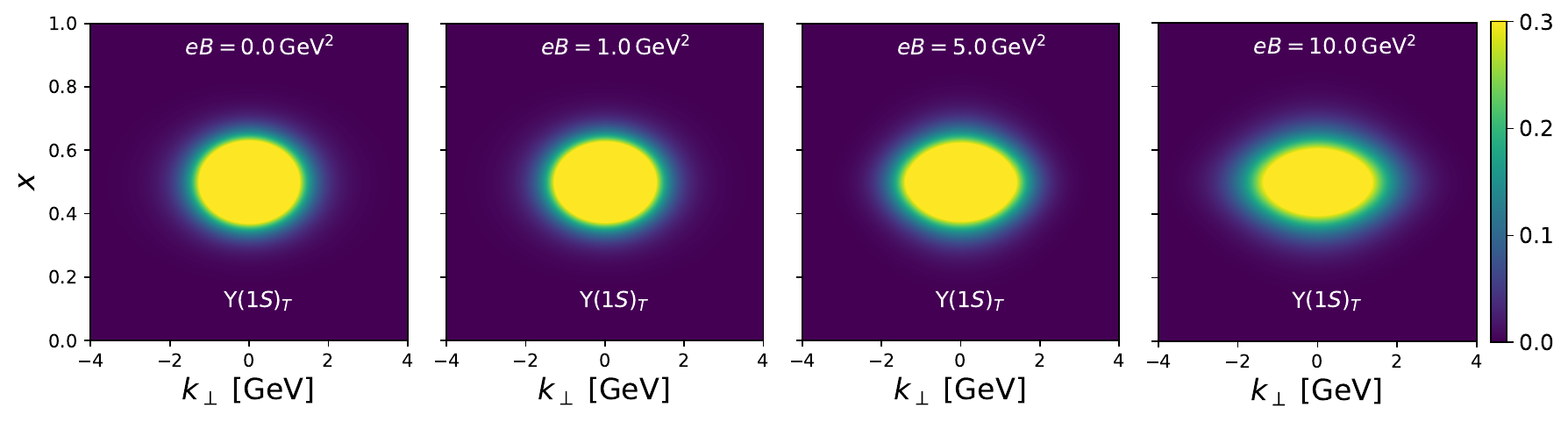}
    \caption{LFWFs of the transverse vector bottomonia in several different magnetic fields:  (Top panels) $\Upsilon(3S)_T$, (Middle panels) $\Upsilon(2S)_T$, and (Bottom panels) $\Upsilon(1S)_T$ states.}
    \label{fig:bottom_LFWF_T}
\end{figure*}

%---------------------------------------------------------

Since the densities are defined in momentum space, an informative quantity is the expectation value of the momentum squared $\expval{k^2}$, which is similar to the root-mean-square radius $\expval{r^2}$ in coordinate space.
In particular, under a magnetic field, it is useful to separate $\expval{k^2}$ into transverse and longitudinal components.
The magnetic-field dependence of $\expval{\tfrac{1}{2}k_\perp^2}$ and $\expval{k_z^2}$ is shown in Fig.~\ref{fig:charm_asym}. Although LFWFs are expressed in terms of the light-front variable $x$, we can still evaluate $\expval{k_z^2}$ via Eq.~\eqref{eq:k_z}. At zero magnetic field, $\expval{\tfrac{1}{2}k_\perp^2} = \expval{k_z^2}$, consistent with the spherical symmetry of the wave function. Under finite magnetic fields, $\expval{\tfrac{1}{2}k_\perp^2}$ increases linearly for both $J/\psi_T$ and $\psi(2S)_T$, while $\expval{k_z^2}$ shows only a slight increase. Interestingly, in the weak-field region, nontrivial behavior emerges due to shape transitions. In particular, for $\psi(2S)_T$, $\expval{k_z^2}$ temporarily exceeds $\expval{\tfrac{1}{2}k_\perp^2}$.
The anisotropy parameter $\epsilon_{\mathrm{LF}}$, plotted in the right panel of Fig.~\ref{fig:charm_asym}, quantifies the deformation of LFWFs. It clearly shows that the wave functions become more elongated in the transverse direction as the magnetic field increases. This parameter provides a useful measure of rotational symmetry breaking induced by the magnetic field and can serve as a point of comparison with other models.
The pseudoscalar and longitudinal vector charmonia exhibit similar trends in their momentum expectation values and anisotropy. However, due to level mixing and state reshuffling, especially in the region of avoided crossings, oscillatory behavior appears in $\expval{k_z^2}$ and $\epsilon_{\mathrm{LF}}$. These features reflect the sensitivity of the excited states to the underlying structural reorganization in the presence of the magnetic field.

Figure~\ref{fig:charm_landau} is a close-up view of the weak-field region of Fig.~\ref{fig:charm_asym}.
One of the interesting features is the shape transition behavior.
The magnetic-field dependence of the transverse momentum expectation value, $\left\langle \tfrac{1}{2}k_\perp^2 \right\rangle$, for transverse charmonia exhibits three different regimes for excited states: (i) a regime where $\left\langle \tfrac{1}{2}k_\perp^2 \right\rangle$ increases due to the elongation of LFWFs (equivalently, the squeezing of spatial wave function) by the Landau level potential, (ii) a regime where the transverse nodes of wave functions gradually disappears, and (iii) a regime where the LFWF is again elongated.
Notably, in the regime~(ii), the shape transition leads to a pronounced decrease in $\left\langle \tfrac{1}{2}k_\perp^2 \right\rangle$, and this effect is more significant for the excited states than for the ground state. 
This behavior reflects the stronger sensitivity of excited states to the confinement dynamics modified by the magnetic field.
We note that this transition behavior can be seen more clearly in bottomonia.

It is also of great importance to examine the projections of the LFWF density onto longitudinal and transverse distributions~\eqref{eq:pdf}, where the $f(x)$ has important meaning in vacuum as a PDF.
Due to the nonrelativistic nature of the quarkonia, the predicted vacuum PDF $f(x)$ from the nonrelativistic quark model~\cite{Wu:2025rto} is comparable with those other models including the light-front Hamiltonian approach~\cite{Lan:2019img}. 

Figure~\ref{fig:charm_pdf} illustrates the $f(x)$ and $f(k_\perp)$ distributions for selected charmonium LFWFs at various magnetic field strengths.
We find that the $f(x)$ distribution is nearly unchanged for the $J/\psi_T$ ground state and the 1st state.
This behavior is expected, as the magnetic-field–dependent potential in the nonrelativistic model does not explicitly involve $z$ or $k_z$ at leading order, whereas the $r_\perp^2$ term does as shown in Eq.~\eqref{eq:Hamiltonian}. 
In contrast, for the $\psi(2S)_T$ state, the three peaks in the $f(x)$ distribution become more pronounced, while the valleys corresponding to the nodal structure deepen. 
The 2nd state also exhibits a transition from a single peak to three peaks due to the avoided crossing. 
These strong modifications of the excited-state $f(x)$ distributions reflect the robustness of their nodal structure.

%---------------------------------------------------------

\begin{figure*}[t]
    \centering
    \includegraphics[width=0.95\textwidth]{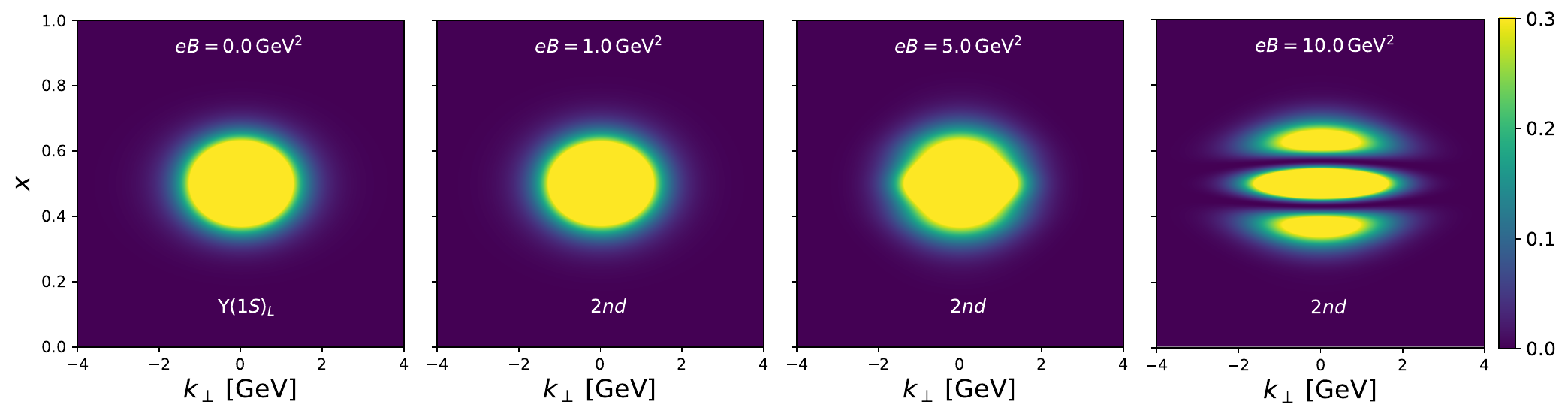}
    \includegraphics[width=0.95\textwidth]{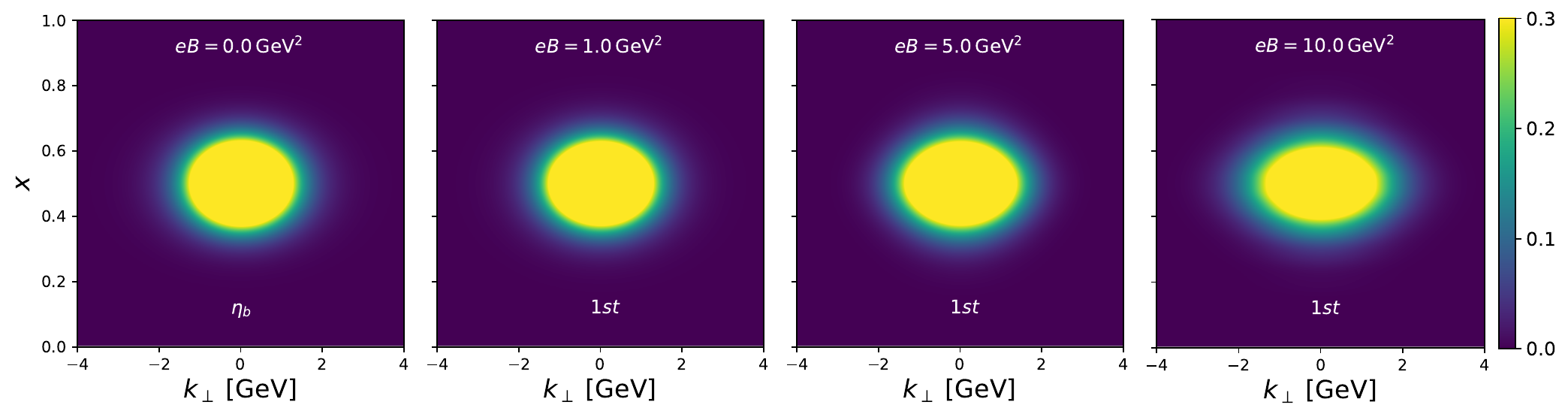}
    \caption{LFWFs of pseudoscalar and longitudinal vector bottomonia in several different magnetic field strengths: (Top panels) the 2nd state and (Bottom panels) the 1st state.}
    \label{fig:bottom_LFWF_L}
\end{figure*}

\begin{figure*}[t]
    \centering   
    \includegraphics[width=0.95\textwidth]{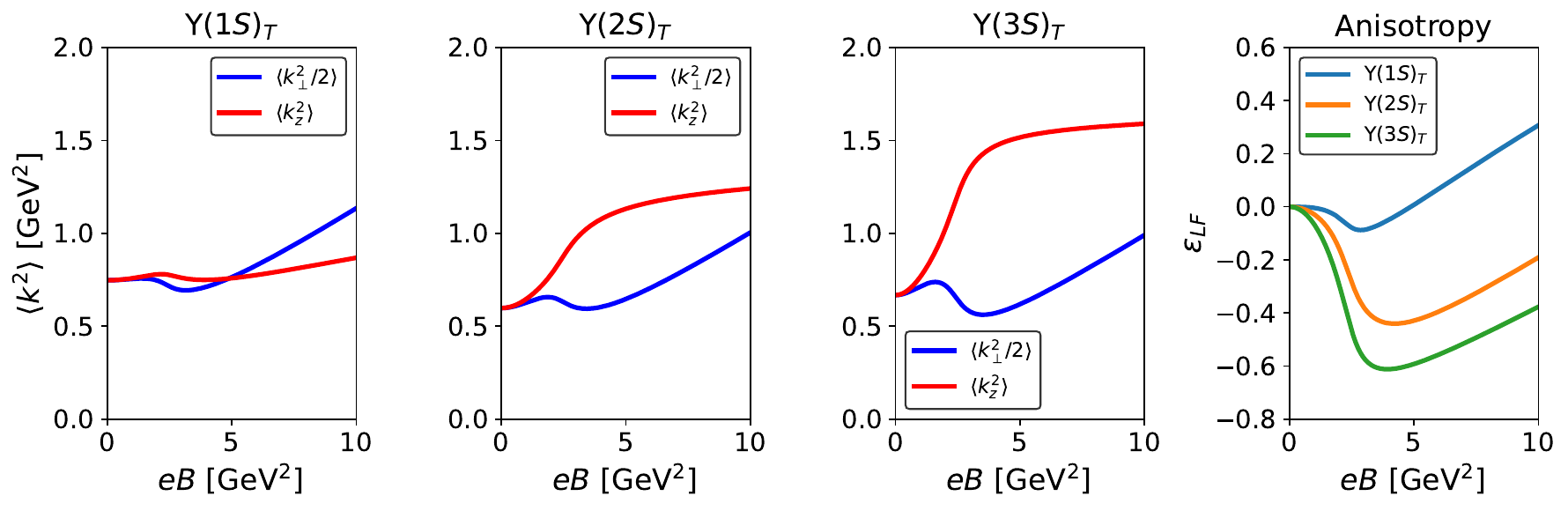}
    \caption{Magnetic field dependence of $\expval{k^2}$ and the anisotropy parameter $\epsilon_{\mathrm{LF}}$ for $\Upsilon(1S)_T, \Upsilon(2S)_T,$ and $\Upsilon(3S)_T$ states. }
    \label{fig:bottom_asym}
\end{figure*}

%---------------------------------------------------------

On the other hand, the $f(k_\perp)$ distribution broadens significantly, as previously mentioned. The distribution extends from $k_\perp = 2$ GeV up to $k_\perp = 4$ GeV, indicating the typical scale of charmonium LFWFs. Moreover, the two-peak structure observed in $\psi(2S)_T$ at low magnetic fields reduces to a single peak at strong fields, marking a clear change in the transverse structure.

Here, we comment on the light-front Hamiltonian approach~\cite{Wen:2025dwy}.
The modification of $f(x)$ distribution is particularly important, because the magnetic-field–dependent potential term in the light-front Hamiltonian is written as the form $r_{\perp i}^2 / x_i$.
This form explicitly depends on $x$, which is clearly different from the potential with $r_{\perp}^2$ in the nonrelativistic Hamiltonian~\eqref{eq:Hamiltonian}.
To explore this further, we will examine the effect of a relativistic correction to the $f(x)$ distribution in Sec.~\ref{sec:rel_cor}.
In addition, in the light-front Hamiltonian approach, the $J/\psi$ density evolves from a single-peak to a two-peak structure resembling a P-wave state. In contrast, the present model shows no such mixing with $\chi_c$ states, since we work in relative coordinates and omit center-of-mass couplings to the magnetic field. 
Identifying the critical magnetic field corresponding to the avoided crossing is essential, for instance, in determining how large the effect of the center-of-mass couplings is, in contrast to the present work.

%---------------------------------------------------------

\begin{figure*}[t]
    \centering
    \includegraphics[width=0.95\textwidth]{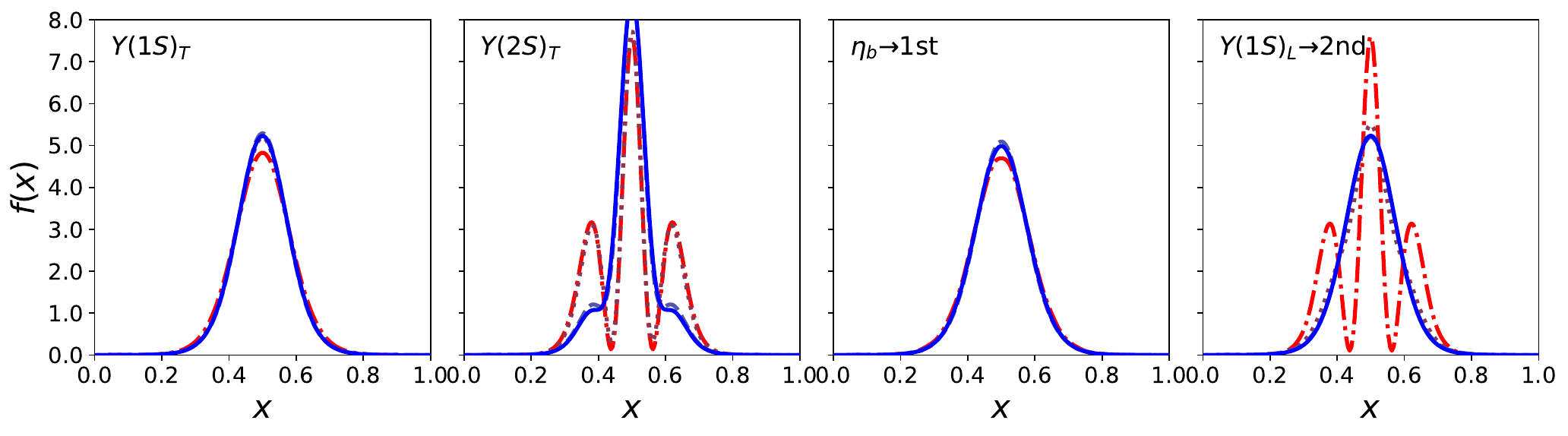}
    \includegraphics[width=0.95\textwidth]{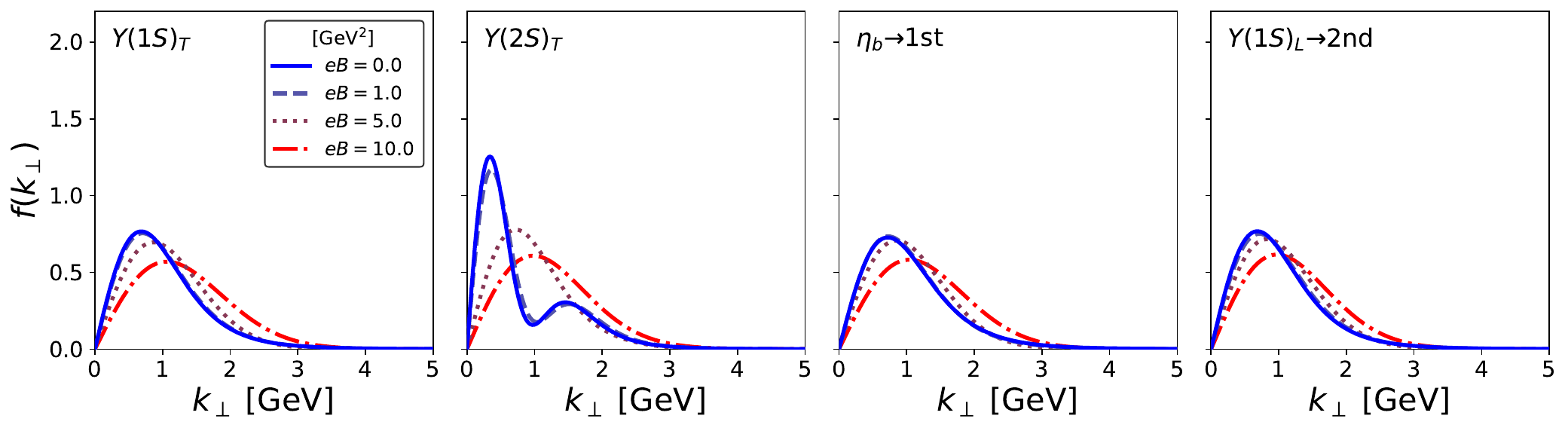}
    \caption{Longitudinal $f(x)$ and transverse $f(k_\perp)$ distributions of selected bottomonia at several magnetic field strengths.}
    \label{fig:bottom_pdf}
\end{figure*}

%---------------------------------------------------------

%---------------------------------------------------------
\subsection{Bottomonia}
%---------------------------------------------------------

Following our detailed investigation of charmonium states, we now turn to the bottomonium sector.
The bottom quark have a heavier mass and a smaller electric charge than the case of the charm quark. These features render bottomonia more nonrelativistic and less sensitive to external magnetic fields compared to charmonia. Nonetheless, they provide an important benchmark to test the universality of magnetic-field effects across heavy quarkonia.

The magnetic field dependence of the bottomonium mass spectra is shown in Fig.~\ref{fig:bottom_mass}, where we recalculated with a denser set of data points compared with previous works~\cite{Yoshida:2016xgm}. 
The transverse vector states [$\Upsilon(1S,2S,3S)_T$] are displayed in the left panel, and the pseudoscalar [$\eta_b(1S,2S,3S)$] and longitudinal vector [$\Upsilon(1S,2S,3S)_L$] states, affected by mixing, are shown in the right panel.
In the right panel, we compute the six lowest eigenstates below the $B\bar{B}$ threshold and label them as the 1st-6th states. This allows us to track the avoided crossings and evolution of the eigenstates in the presence of mixing induced by the $-\bm{\mu}_i\cdot\bm{B}$ interaction.

%---------------------------------------------------------

\begin{figure}[b]
    \centering
    \includegraphics[width=0.45\textwidth]{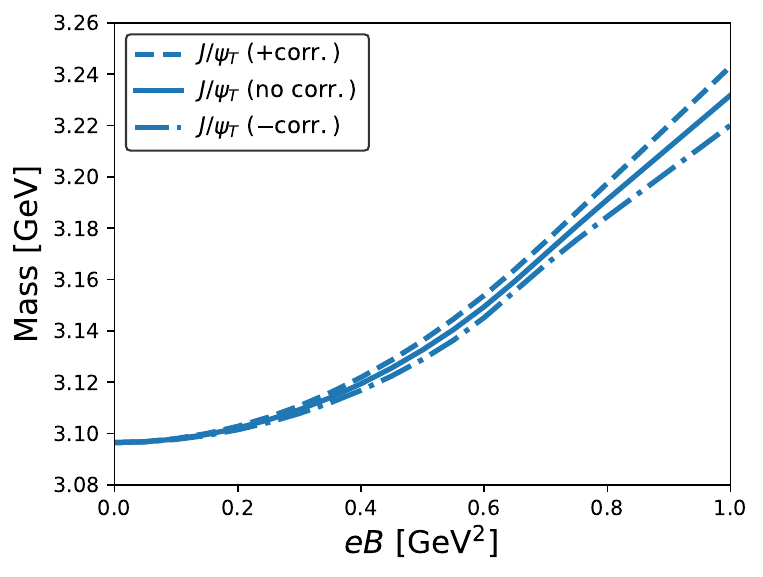}
    \caption{Magnetic-field dependence of the mass spectrum for $J/\psi_T$ with and without the $k_z^2$ correction in the Landau level term~\eqref{eq:correction}. The correction with a positive (negative) sign slightly increases (decreases) the mass.}
    \label{fig:charm_mass_T_rc}
\end{figure}

\begin{figure*}[t]
    \centering
    \includegraphics[width=0.80\textwidth]{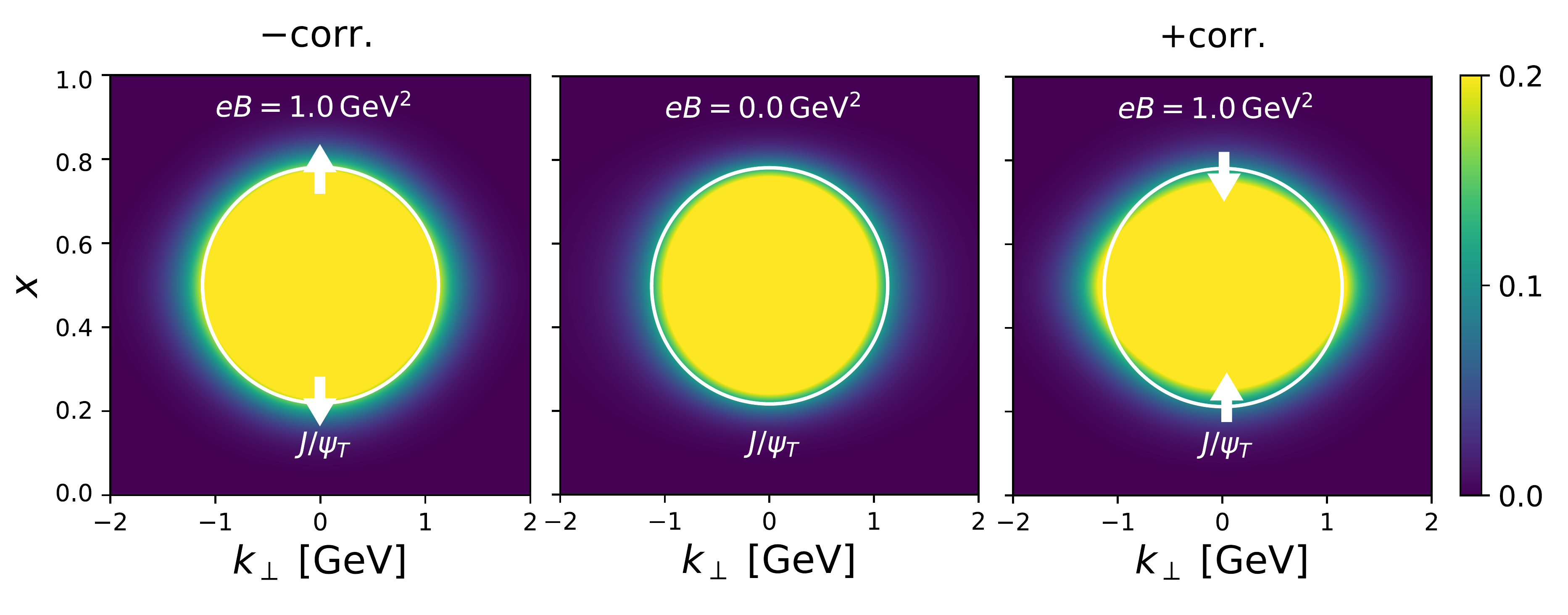}
    \caption{LFWFs of the $J/\psi_T$ state at $eB=0$ and $1.0$ GeV$^2$ with a positive or negative correction~\eqref{eq:correction}. The longitudinal momentum fraction distribution with a positive (negative) correction becomes more squeezed (elongated).}
    \label{fig:charm_LFWF_T1_rc}
\end{figure*}

\begin{figure}[t]
    \centering
    \includegraphics[width=0.42\textwidth]{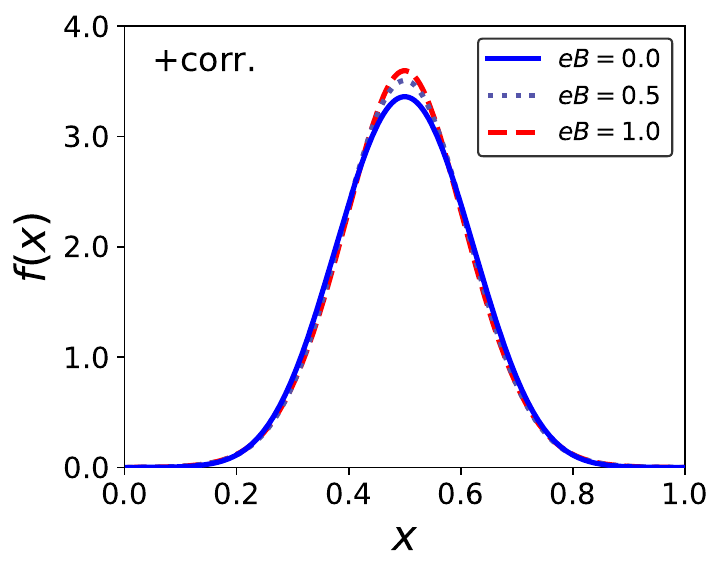}
    \includegraphics[width=0.42\textwidth]{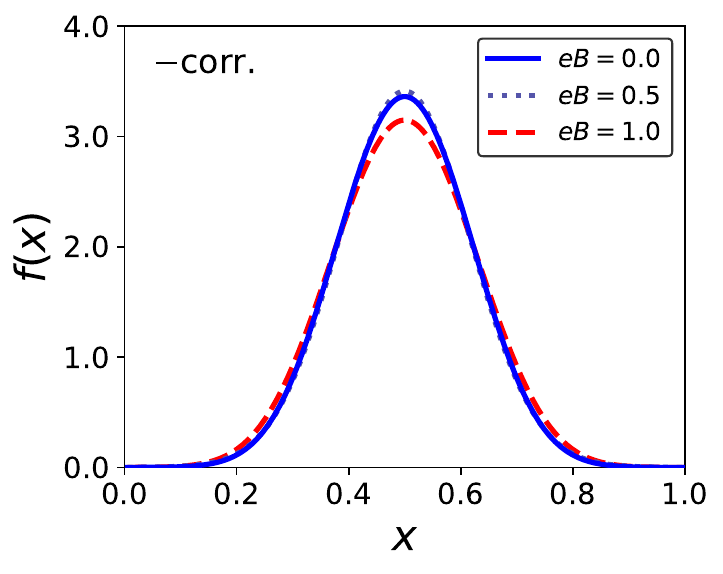}
    \caption{PDFs of the $J/\psi_T$ states at various magnetic field strengths with a positive or negative correction~\eqref{eq:correction} included. Up to $eB$=1 GeV$^2$, the PDFs with the positive (negative) correction become slightly narrower (broader) as the magnetic field increases.}
    \label{fig:charm_pdf_rc}
\end{figure}

%---------------------------------------------------------

As expected, the transverse states ($\Upsilon_T$) show linear mass increases with increasing $eB$, dominated by the quark Landau level effect. The mass shift becomes slightly more pronounced for higher excited states due to their larger transverse size of wave functions.
Compared to charmonium, however, the absolute values of shifts are smaller due to the suppression from the larger $b$-quark mass and smaller charge ($|e_b| = e/3$).

The right panel of Fig.~\ref{fig:bottom_mass} shows the longitudinal sector.
In the weak field, we find level repulsions between the $\eta_b(nS)$ and $\Upsilon(nS)_L$ states.
The avoided crossings, such as 2nd-3rd and 3rd-4th, occur at slightly higher field strengths compared to charmonium due to the weaker mixing.
These crossings are visible across the wide range of magnetic field strengths.

The LFWF densities for the transverse vector bottomonium states are plotted in Fig.~\ref{fig:bottom_LFWF_T} for $\Upsilon(1S,2S,3S)_T$ across several magnetic field strengths. 
As in the charmonium case, the densities are shown in momentum space and display transverse broadening due to the magnetic-field-dependent confining potential, which scales as $B^2 r_\perp^2$ and is weighted by $q_b^2 / m_b$, thus making the effect less pronounced.
The nodal structures of the excited states change into multiple oblate peaks with increasing magnetic field.

The corresponding LFWF densities for the mixed pseudoscalar and longitudinal vector bottomonia (first and second states) are shown in Fig.~\ref{fig:bottom_LFWF_L}, where we display only the lowest two states for examples.
Their evolution with increasing magnetic field already reveals key features of the magnetic-induced mixing. 
In particular, the second state develops a three-peak structure at $eB=10 \,\mathrm{GeV}^2$, indicating that the energy of the radial excitation falls below that of the nodeless wave function.
This reflects a characteristic reshuffling of state composition: as the magnetic field increases, the eigenstates continuously exchange their internal structure, such as nodal patterns and dominant components. 
For instance, the third state can be tracked from the mass spectrum evolution as transitioning from a primarily 2S-like state at low $eB$, becoming 1S-like near the avoided crossing, and eventually acquiring 3S-like features at higher $eB$. 
Such mixing leads to nontrivial deformation patterns in the LFWFs and illustrates how magnetic fields  reorganize the internal structure of heavy quarkonia.

To quantify the deformation in momentum space, we compute $\expval{\frac{1}{2}k_\perp^2}$ and $\expval{k_z^2}$ in Fig.~\ref{fig:bottom_asym} for $\Upsilon(1S,2S,3S)_T$. 
We find that the transverse momentum for all three states first increases, then decreases in the weak-field region, and again increases around $eB = 3$ GeV$^2$ as previously discussed for charmonia in Fig.~\ref{fig:charm_landau}. 
On the other hand, the longitudinal momentum rises rapidly at first, but its increase becomes more gradual as $eB$ continues to grow.
Interestingly, the anisotropy parameter $\epsilon_{\mathrm{LF}}$ shows a negative value up to $eB=10$ GeV$^2$ for the excited states, indicating that the longitudinal momentum always takes a larger value. 
This is due to the robustness of their nodal structure.
Compared to the charmonium system, it is clear that the deformation is less extreme due to weaker magnetic responses. 
This provides further evidence of the robustness of the bottomonium system.

Lastly, the longitudinal $f(x)$ and transverse $f(k_\perp)$ distributions for bottomonia at various magnetic field strengths are shown in Fig.~\ref{fig:bottom_pdf}. 
For the ground state $\Upsilon(1S)_T$, the $f(x)$ distribution remains largely unaffected by the magnetic field. 
For excited states such as $\Upsilon(2S)_T$, the $f(x)$ distribution becomes narrower in $x$, consistent with the increase of $\expval{k_z^2}$, and the valleys become deepen and the peaks of the shoulders enhanced.
The 2nd state in the longitudinal sector also display evolving multi-peak structures due to nodal features and mixing-induced reshaping.
The $f(k_\perp)$ distribution extends up to around 3 GeV and stay the same, but the width of the peak is broaden.
The two-peak structure for $\Upsilon(2S)_T$ is reduced to a one-peak structure due to the disappearance of the transverse nodes.

%=========================================================
\subsection{Relativistic correction} 
\label{sec:rel_cor}
%=========================================================

So far, we have demonstrated the deformation of the LFWFs under magnetic field using the nonrelativistic Hamiltonian model as input. 
One of the dominant effects observed is transverse momentum broadening. 
However, within this setup, the longitudinal distribution remains largely unmodified for the ground states as seen in Figs.~\ref{fig:charm_pdf} and \ref{fig:bottom_pdf}.

The recent result based on the light-front Hamiltonian approach~\cite{Wen:2025dwy} has shown that the PDFs can also be significantly altered.
However, in the nonrelativistic Hamiltonian~\eqref{eq:Hamiltonian}, since the magnetic-field-dependent potential (i.e., the Landau level term) is proportional to $r_\perp^2$, this term leads to the transverse momentum broadening but does not directly contribute to the longitudinal momentum distribution.
In this sense, while our numerical results are reasonable, it might be important to discuss possible scenarios of the PDF modifications within our model.

In this work, we phenomenologically introduce a $k_z$-dependent Landau level term as a relativistic correction (i.e., a higher-order term in the $1/m$ expansion):
\begin{equation}\label{eq:correction}
V^{\pm}_L = \frac{q^2B^2}{8\mu}r_\perp^2 \left(1 \pm \frac{k_z^2}{2\mu^2}\right).
\end{equation}
This term is not exactly derived from the two-body Hamiltonian, but a possible origin from a one-body Hamiltonian is explained in Appendix~\ref{app:correction}.
Although the sign of this correction is suggested to be negative in Appendix~\ref{app:correction}, we treat the sign as a free parameter to explore its impact on the PDFs. 
We then diagonalize the Hamiltonian with the correction term~\eqref{eq:correction} and focus on only the $J/\psi_T$ as a simple example where the pseudoscalar-vector meson mixing is absent.
We note that the correction may be valid only for values of $\sqrt{qB}$ smaller than the quark mass because of the $\sqrt{qB}/m$ expansion.

%---------------------------------------------------------
\begin{figure}[t]
    \centering
    \includegraphics[width=0.42\textwidth]{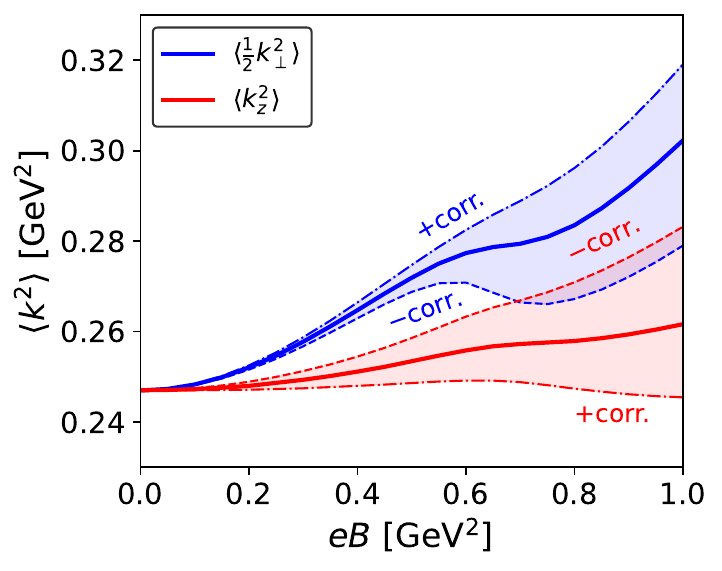}
    \caption{Magnetic-field dependence of the momentum expectation values, $\expval{\frac{1}{2}k_\perp^2}$  and $\expval{k_z^2}$ for the $J/\psi_T$ state with positive and negative corrections. The correction term~\eqref{eq:correction} significantly modifies $\expval{k_z^2}$ with increasing magnetic field. }
    \label{fig:charm_asym_rc}
\end{figure}
%---------------------------------------------------------

The impact of the additional $k_z^2$ term~\eqref{eq:correction} on the mass spectrum is illustrated in Fig.~\ref{fig:charm_mass_T_rc}. 
While this term slightly increases or decreases the mass of the $J/\psi_T$ depending on the sign, the linear trajectory remains unaffected. 
The corresponding LFWFs, shown in Fig.~\ref{fig:charm_LFWF_T1_rc}, with a positive (negative) correction are now more squeezed (elongated) in the longitudinal direction. 
One important implication is that the PDFs become slightly narrower or broader, as illustrated in Fig.~\ref{fig:charm_pdf_rc}. For comparison, in the light-front Hamiltonian approach~\cite{Wen:2025dwy}, the PDFs for charmonium states corresponding to $\eta_c \to$ 1st and $J/\psi_T$\footnote{In Ref.~\cite{Wen:2025dwy}, these states are labeled by a special quantum number $m_{PC} = -1$ and $+1$.} at $eB = 1\, \text{GeV}^2$ are slightly broader and narrower, respectively (see Fig.~9 of Ref.~\cite{Wen:2025dwy}).
These behaviors may reflect complex relativistic dynamics but might be roughly interpreted as the $k_z^2$ correction~\eqref{eq:correction}.
Furthermore, Fig.~\ref{fig:charm_asym_rc} shows the magnetic-field dependence of $\expval{\tfrac{1}{2}k_\perp^2}$ and $\expval{k_z^2}$. 
Notably, the inclusion of the $k_z^2$ correction with a plus sign reverses the behavior of $\expval{k_z^2}$, changing from an increasing to a decreasing trend with growing magnetic field strength. 
This reversal is directly reflected in the modification of the PDFs.

%=========================================================
\section{Conclusion and Outlook} 
%=========================================================
\label{sec:conclusion}

In this work, we have investigated the structural modifications of heavy quarkonia, charmonia and bottomonia, under strong magnetic fields in the constituent quark model. 
By solving the two-body Schrödinger equation with the cylindrical Gaussian expansion method~\cite{Suzuki:2016kcs, Yoshida:2016xgm}, 
we obtained precise wave functions that respect the cylindrical symmetry imposed by the magnetic field. 
Here, these wave functions were then transformed into LFWFs to reveal their momentum-space structure. 

Our results show that the presence of a strong magnetic field leads to a clear transverse momentum broadening of the LFWFs, driven by the magnetic-field–induced potential in the nonrelativistic Hamiltonian~\cite{Suzuki:2016kcs, Yoshida:2016xgm}. In contrast, the longitudinal momentum fraction distribution for the ground states remains largely unchanged within this framework. For excited states, however, we find characteristic shape transitions and nodal restructuring near avoided crossings, which result in significant modifications to the longitudinal momentum distributions. 
Notably, these avoided crossings play a key role in reorganizing the level structure and internal dynamics of quarkonia under external magnetic fields.
To explore the effect beyond this framework, we have introduced a phenomenological relativistic correction to the Landau level term that significantly modifies the longitudinal structure.
Furthermore, we also expect that bottomonia behave more nonrelativistically and are less sensitive to external magnetic fields compared to charmonia. 
It is worth noting that the anisotropy parameter may be useful in discussing the structural modifications.
This can be compared with relativistic approaches to further explore the role of relativistic effects.

These findings provide a qualitative understanding of how quarkonia structure is modified by magnetic fields within a nonrelativistic Hamiltonian framework~\cite{Suzuki:2016kcs, Yoshida:2016xgm} and help bridge this understanding with results obtained from the light-front Hamiltonian approach~\cite{Wen:2025dwy}. The insights gained here can guide future lattice QCD studies aimed at quantifying magnetic-field effects on hadron structure.
Further extensions of this work could include anisotropy of the confining or Coulomb potential observed in lattice QCD simulations~\cite{Bonati:2014ksa,Bonati:2016kxj,DElia:2021tfb} and explicit time-dependent magnetic fields~\cite{Guo:2015nsa,Suzuki:2016fof,Dutta:2017pya,Hoelck:2017dby,Bagchi:2018olp,Iwasaki:2021kms} to better approximate the conditions of heavy-ion collisions. 
Additional effects and phenomena related to quarkonia in strong magnetic fields also remain to be explored for future studies. 

% =========================================================
\section*{Acknowledgment}
% =========================================================

A. J. A was partly supported by the RCNP Collaboration Research Network program as the project number COREnet 057.
This work was supported by the Japan Society for the Promotion of Science (JSPS) KAKENHI (Grant No. JP24K07034).

\appendix
%---------------------------------------------------------
\section{Relativistic corrections}
\label{app:correction}
%---------------------------------------------------------

In the main text, we have investigated the impact of a relativistic correction~\eqref{eq:correction}.
In this Appendix, we explain the origin of this term.

From the Dirac equation in magnetic fields, we can derive the single-particle energy~\cite{Rabi:1928,Johnson:1949} [also see, e.g., Eq.~(83) of Ref.~\cite{Hattori:2023egw}]
\begin{equation}
    \epsilon_n = \pm \sqrt{p_z^2 + 2n|qB| + m^2 },
\end{equation}
where the transverse momentum $p_\perp$ is quantized, and $n=0,1,2,\cdots$ is the index of Landau levels.
Now, we consider a nonrelativistic expansion of the single-particle energy as\footnote{This expansion holds in the condition of $\sqrt{p_z^2+2n|qB|}\ll m$.
If $n|qB|$ is large enough, this expansion cannot be used, but it is valid as long as we focus on the lower Landau levels and weak magnetic fields.}
\begin{align}
  \epsilon_n &= m \sqrt{ 1 + \frac{p_z^2+2n|qB|}{m^2} }  \\
             &= m + \frac{p_z^2+2n|qB|}{2m} -\frac{(p_z^2+2n|qB|)^2}{8m^3}+ \dots .
\end{align}
In the last form, the first and second terms are consistent with the mass, the kinetic energy, and the nonrelativistic Landau levels with the cyclotron frequency $|qB|/m$, which are used in the usual nonrelativistic framework. 
On the other hand, the third term is the higher-order corrections proportional to $1/m^3$, which
includes the cross term between $p_z^2$ and $2n|qB|$.
Thus, the momentum $p_z$ and the Landau level effect can be correlated as a higher-order relativistic correction.
The energy shift from this term is
\begin{align}
  \Delta\epsilon_n = -\frac{4n|qB|p_z^2}{8m^3} = - \frac{p_z^2}{2m^2}\times\frac{n|qB|}{m}.
\end{align}
The factor of $- p_z^2/2m^2$ may be regarded as a correction to the original energy shift of Landau levels.

However, the rigorous derivation of the two-body potential from such a single-particle energy is technically difficult.
Because of that, the study toward such a direction is limited so far.
Under such a situation, to reproduce the narrowing of PDF (found in Ref.~\cite{Wen:2025dwy}), we have phenomenologically introduced the relativistic correction~\eqref{eq:correction}.
Note that we can also consider the corrections proportional to $p_z^4/m^3$ or $|qB|^2/m^3$, but they are not relevant for our purpose.

%---------------------------------------------------------
\section{Potential matrix elements}
\label{app:matrix_element}
%---------------------------------------------------------

While the potential matrix elements have been shown in Ref.~\cite{Yoshida:2016xgm},
here we present the analytical expressions for the matrix elements relevant to the Landau-level potential and its relativistic correction.
Using the cylindrical Gaussian basis $\phi_n^{CG}$, the matrix element, relevant for the Landau term, is given by
\begin{align}
\Braket{r_\perp^2} &=\mel{\phi_n^{CG}}{r_\perp^2}{\phi_m^{CG}}\nonumber\\
&= \frac{N_n N_m \pi^{3/2}}{\beta_{nm}^2 \sqrt{\gamma_{nm}}},
\end{align}
where $\beta_{nm} = \beta_n + \beta_m, \gamma_{nm} = \gamma_n + \gamma_m$, and
the normalization factor
\begin{eqnarray}
   N_n = \frac{2^{3/4}\gamma_n^{1/4}\sqrt{\beta_n}}{\pi^{3/4}}.
\end{eqnarray}
The correction term~\eqref{eq:correction}, proportional to $r_\perp^2 k_z^2$, can be computed as
\begin{align}
\Braket{r_\perp^2 k_z^2} &= -\mel{\phi_n^{CG}}{r_\perp^2 \frac{\partial^2}{\partial z^2}}{\phi_m^{CG}}\nonumber\\
&= \frac{2\gamma_m \gamma_n}{\gamma_{nm}}
\left[ \frac{N_n N_m \pi^{3/2}}{\beta_{nm}^2 \sqrt{\gamma_{nm}}} \right],
\end{align}
where $k_z^2$ corresponds to $-\partial^2/\partial z^2$ in coordinate space.

%---------------------------------------------------------
\bibliography{references}
%---------------------------------------------------------

\end{document}